\documentclass[ejs,preprint,twoside,linksfromyear,nameyear,natbib]{imsart}
\usepackage{amsthm,amsmath,amssymb}
\usepackage{hyperref}
\usepackage{breakurl}
\usepackage{graphicx}

\doi{10.1214/15-EJS1092}
\pubyear{2015}
\volume{9}
\issue{2}
\firstpage{2706}
\lastpage{2731}

\startlocaldefs
\endlocaldefs

\begin{document}

\begin{frontmatter}
\title{Improving the INLA approach for approximate Bayesian inference
for latent Gaussian models}
\runtitle{Improving the INLA approach}


%

%
\begin{aug}
\author{\fnms{Egil} \snm{Ferkingstad}\corref{}\ead
[label=e1]{egil.ferkingstad@gmail.com}}
\address{Department of Mathematical Sciences\\
Norwegian University of Science and Technology\\
Trondheim, Norway\\
and\\
Science Institute\\
University of Iceland\\
Reykjavik, Iceland\\
\printead{e1}}
\end{aug}
\medskip\textbf{\and}
\begin{aug}
\author{\fnms{H{\aa}vard} \snm{Rue}\ead[label=e2]{hrue@math.ntnu.no}}
\address{Department of Mathematical Sciences\\
Norwegian University of Science and Technology\\
Trondheim, Norway\\
\printead{e2}}
\end{aug}

\runauthor{E. Ferkingstad and H. Rue}

\begin{abstract}
We introduce a new copula-based correction for generalized linear mixed
models (GLMMs) within the integrated nested Laplace approximation
(INLA) approach for approximate Bayesian inference for latent Gaussian models.
While INLA is usually very accurate, some (rather extreme) cases of
GLMMs with e.g.~binomial or Poisson data have been seen to be
problematic. Inaccuracies can occur when there is a very low degree of
smoothing or ``borrowing strength'' within the model, and we have
therefore developed a correction aiming to push the boundaries of the
applicability of INLA.
Our new correction has been implemented as part of the R-INLA package,
and adds only negligible computational cost. Empirical evaluations on
both real and simulated data indicate that the method works well.
\end{abstract}



%
\received{\smonth{4} \syear{2015}}

\end{frontmatter}
\maketitle

\section{Introduction}

Integrated Nested Laplace Approximations (INLA) were introduced
by~\citet{rue2009} as a tool to do approximate Bayesian inference in
latent Gaussian models (LGMs). The class of LGMs covers a large part of
models used today, and the INLA approach has been shown to be very
accurate\vadjust{\eject} and extremely fast in most cases. Software is provided through
the R-INLA package, see \url{http://www.r-inla.org}.

An important subclass of LGMs is the rich family of generalized linear
mixed models (GLMMs) with Gaussian priors on fixed and random
effects~\citep{breslow1993,mcculloch2008}. The use of INLA for Bayesian
inference for GLMMs was investigated by~\citet{fong2010}, who
reanalyzed all of the examples from~\citet{breslow1993}. \citet
{fong2010} found that INLA works very well in most cases, but one of
their examples shows some inaccuracy for binary data with few or no
replications. In this paper, we introduce a new correction term for
INLA, significantly improving accuracy while adding negligibly to the
overall computational cost.

%
\begin{figure}[t]
\includegraphics[width=\textwidth]{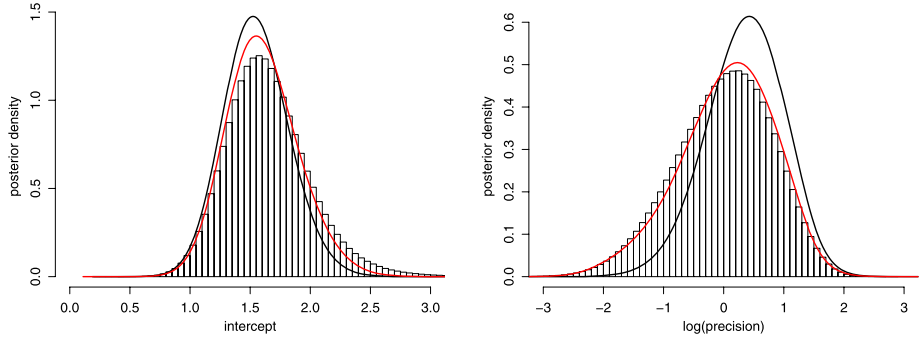}
\caption{Minimal example defined in Equation~\eqref{eq:minimal}. The
histograms show posterior distributions from a long MCMC run (ten
chains of one million iterations each), the black curves show the
posteriors from INLA, while the red curves show the posteriors using
our new correction to INLA.}
\label{figure1}
\end{figure}

To set the scene, we consider a minimal simulated example illustrating
the problem (postponing more thorough empirical evaluations until
Section~\ref{sec:examples}). Consider the following model: For
$i=1,2,\ldots,n$, let $\mathrm{Prob}(y_i=0)=1-p_i$, $\mathrm
{Prob}(y_i=1)=p_i$, and
%
\begin{equation}
\label{eq:minimal}
\mathrm{logit}(p_i) = \beta+ u_i,
\end{equation}
where $u_i \sim N(0,\sigma^2)$, iid. Let the precision $\sigma^{-2}$
have a $\mathrm{Gamma(1,1)}$ prior, while the prior for $\beta$ is $N(0,1)$.
We simulated data from this model, setting $n=100$, $\sigma^2=1$ and
$\beta=2$. Figure~\ref{figure1} shows the resulting posterior
distributions for the intercept $\beta$ and for the log precision, $\log
(\sigma^{-2})$, where the histograms show results from long MCMC runs
using JAGS~\citep{plummer2013}, the black curves show posteriors from
INLA without any correction, and the red curves show results using the
new correction defined in Section~\ref{sec:method}. While some of our
later examples show more dramatic differences between INLA and long
MCMC runs, these results exemplify quite well our general experience
with using INLA for ``difficult'' binary response GLMMs: Variances of
both random and fixed effects tends to be underestimated, while the
means of the fixed effects are reasonably well estimated.

%
\begin{figure}[t!]
\includegraphics[width=\textwidth]{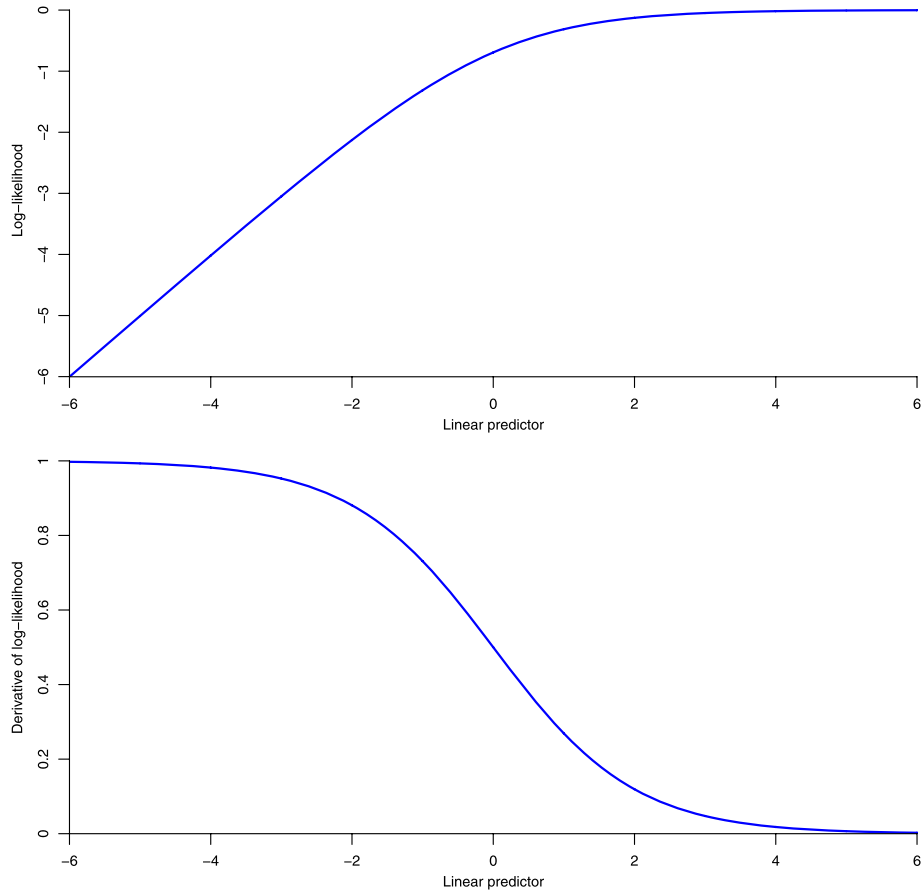}
\caption{Log-likelihood (top panel) and derivative of log-likelihood
(bottom panel) for a single Bernoulli observation as a function of the
linear predictor in a logistic model.}
\label{figure2}
\end{figure}

One part of the problem is that the usual assumptions ensuring
asymptotic validity of the Laplace approximation do not hold here (for
details on asymptotic results, see the discussion in Section 4 of \citet
{rue2009}). The independence of the random effects make the
effective number of parameters~\citep{spiegelhalter2002}
on the order of the number of data points. In more complex models,
there is often some amount of smoothing or replication that alleviates
the problem, but it may still occur. Except in the case of spline
smoothing models~\citep{kauermann2009}, there is a lack of strong
asymptotic results for random effects models with a large effective
number of parameters. In the simulation from model~\eqref{eq:minimal},
the data provide little information about the parameters, with the
shape of the likelihood function adding to the problem.
Figure~\ref{figure2} illustrates the general problem. Here, the top
panel shows the log-likelihood of a single Bernoulli observation $Y$ as
a function of the linear predictor $\eta$, i.e.~$\log(\text
{Prob}(Y=1))=\log(p)$ where $\text{logit}(p)=\eta$. The bottom panel
shows the corresponding derivative. We see that the log-likelihood gets
very flat (and the derivative near zero) for high values of $\eta$, so
inference will get difficult.

Bayesian and frequentist estimation for GLMMs with binary outcomes has
been given some attention in the recent literature~\citep
{capanu2013,grilli2014,sauter2015}, but a computationally efficient
Bayesian solution appropriate for the INLA approach has been lacking.
An alternative to our new approach would be to consider higher-order
Laplace approximations~\citep{shun1995,raudenbush2000,evangelou2011},
other modifications to the Laplace approximation~\citep
{ruli2015,ogden2015}, or expectation propagation-type solutions~\citep
{cseke2010}, but we view them as too costly to be applicable for
general use in INLA.
The motivation for using INLA is speed, so we see it as a design
requirement for any correction that it should add minimally to the
running time of the algorithm.

We proceed as follows.
In Section~\ref{sec:method}, we present a derivation of our new
correction method. Section~\ref{sec:examples} presents empirical
studies, both on real and simulated data, showing that the method works
well in practice. Finally Section~\ref{sec:conclusion} gives a brief
discussion and some concluding remarks.

\section{Methodology}
\label{sec:method}

Consider a latent Gaussian model \citep{rue2009}, with
hyperparameters $\theta=(\theta_1,\ldots,\theta_p)'$, latent field
$x=(x_1,\ldots,x_n)'$ and observed data
$y=\{y_i: i \in\mathcal{I}\}$ (for $\mathcal I \subseteq\{1,\ldots,n\}$),
where the joint distribution may be written as
\[
\pi({x},{\theta},{y})=\pi({\theta})\pi({x}|{\theta})\prod_{i \in
\mathcal{I}} \pi(y_i|x_i,{\theta})
\]
where $\pi({x}|{\theta})$ is a multivariate Gaussian density. We want
to approximate the
posterior marginals $\pi(x_i|y)$ and $\pi(\theta_j|y)$. The Laplace
approximation of $\tilde\pi(\theta| y)$ is
%
\begin{equation}
\label{eq:laplace}
\tilde\pi(\theta| y) \propto\left.\frac{\pi(x,\theta,y)}{\tilde\pi
_G(x|\theta,y)}\right\vert_{ x= \mu(\theta)}
\end{equation}
where $\tilde\pi_G({x}|{\theta},{y})$ is a Gaussian approximation
found by matching the mode and the curvature at the mode of $\pi
(x|\theta,y)$, and $\mu(\theta)$ is the mean of the Gaussian approximation.

Given $\tilde\pi(\theta| y)$ and some approximation $\tilde\pi
(x_i|\theta,y)$ (see below), the posterior marginals of interest are
calculated using numerical integration:
%
\begin{eqnarray}
\tilde\pi(\theta_j|y) &=& \int\tilde\pi(\theta| y) \mathrm{d}\theta
_{-j},\nonumber\\
\tilde\pi(x_i|y) &=& \int\tilde\pi(x_i|\theta,y) \tilde\pi(\theta|
y) \mathrm{d}\theta.
\label{eq:integral2}
\end{eqnarray}

In the current implementation of INLA, the $\tilde\pi(x_i|\theta,y)$
used in~\eqref{eq:integral2} are approximated using skew normal
densities $\tilde\pi_\text{SN}(x_i|\theta,y)$~\citep{azzalini1999},
based on a second Laplace approximation; see Section 3.2.3 of~\citet
{rue2009} for details.
Notice that, in Equation~\eqref{eq:laplace} we use a Gaussian
approximation $\tilde\pi_G({x}|{\theta},{y})$, with marginals
$\tilde\pi_G({x_i}|{\theta},{y}), \ i=1,\ldots,n$. Thus, both $\pi
_\text{SN}(x_i|\theta,y)$ and $\tilde\pi_G({x}|{\theta},{y})$ are
approximations to the marginals $\pi(x_i|\theta,y)$, but the $\pi_\text
{SN}(x_i|\theta,y)$ are more accurate since they are based on a second
Laplace approximation. In~\eqref{eq:laplace} we need to approximate the
full joint distribution $\pi(x|\theta,y)$.
Our basic idea is to use the improved approximations $\tilde\pi
_{SN}(x_i | \theta,y)$ in order to construct a better approximation to
the joint distribution $\pi(x|\theta,y)$.
We aim for an approximation of $\pi(x|\theta,y)$ that retains the
dependence structure of the Gaussian approximation $\tilde\pi
_G(x|\theta,y)$, while having the improved marginals $\tilde\pi
_{SN}(x_i | \theta,y)$.
This can be achieved by using a Gaussian copula.

Before we describe the copula construction, we need to define some
notation. First, for $i = 1,\ldots,n$, let
$\tilde\mu_i(\theta)$ and $\tilde\sigma_i^2(\theta)$ denote the mean
and variance of each marginal $\tilde\pi_{SN}(x_i | \theta,y)$, and let
$F_i$ be the cumulative distribution function corresponding to $\tilde
\pi_{SN}(x_i | \theta,y)$.
Second, let $\tilde x_i \sim F_i$ and assume that $\tilde F_i$ is the
distribution of $\tilde z_i = (\tilde x_i - \tilde\mu_i(\theta))/\sigma
_i(\theta)$.
As usual, $\Phi$ denotes the cumulative standard Gaussian distribution
function. Furthermore, let $\mu_i(\theta)$ and $\sigma_i^2(\theta)$
denote the marginal means and variances of the Gaussian approximation
$\tilde\pi_G({x}|{\theta},{y})$,
let $Q(\theta)$ be the precision matrix of $\tilde\pi_G({x}|{\theta
},{y})$, and
let $x=(x_1,\ldots,x_n)$ where $x \sim\tilde\pi_G({x}|{\theta},{y})$,
and define $z_i = (x_i - \mu_i(\theta))/\sigma_i(\theta), \ i=1,\ldots,n$.
Note that we have $\tilde\sigma_i^2(\theta) \equiv \sigma_i^2(\theta
)$ from the definition of $\tilde\pi_{SN}(x_i | \theta,y)$ (the
construction of the skew normal changes the mean and adds skewness, but
keeps the variance unchanged; again, see Section 3.2.3 of~\citet
{rue2009} for detailed explanations), so from here on we denote both
simply by $\sigma_i^2(\theta)$.

We will now show how to construct a joint distribution having marginals
$F_i$ and the dependence structure from $\tilde\pi_G(x|\theta,y)$,
using a Gaussian copula (see e.g.~\citet{nelsen2007} for a general
introduction). First, note that $\Phi(z_i) \sim\text{U}[0,1]$ by the
probability integral transform (PIT).
Let $\tilde z_i = \tilde F_i^{-1}(\Phi(z_i))$.
Applying the inverse of the PIT then yields that $\tilde z_i \sim
\tilde F_i$, from which it follows that
$\tilde x_i = \sigma_i(\theta) \tilde z_i + \tilde\mu_i(\theta)$ is
distributed as $\tilde x_i \sim F_i$, which is the marginal
distribution we want. Since we have only done marginal transformations,
the dependence structure of the original $x = (x_1, \ldots x_n)'$ is
still intact.
Thus, to construct the new approximation to the joint distribution
$\tilde\pi(x | \theta,y)$, we define the transformed value $\tilde
x_i$ as follows:
%
\begin{equation}
\tilde x_i = \sigma_i(\theta) \tilde F_i^{-1} \left[ \Phi\left(\frac
{x_i-\mu_i(\theta)}{\sigma_i(\theta)}\right)\right] + \tilde\mu
_i(\theta).
\label{eq:copula}
\end{equation}

We may simplify the construction above by replacing the $\tilde F_i$
in~\eqref{eq:copula} by $\Phi$.
This means that we do not correct for skewness, but we take advantage
of the improved mean $\tilde\mu_i(\theta)$ from $\tilde\pi_{SN}(x_i |
\theta,y)$. We denote this as the ``mean only'' correction. (We will
later discuss the possibility of retaining $\tilde F_i$ as a skew
normal; this we denote as the ``mean and skewness'' correction.)
In the simple ``mean only'' case, the transformation reduces to a shift
in mean:
\[
\tilde x_i= x_i - \mu_i(\theta) + \tilde\mu_i(\theta),
\]
the Jacobian is equal to one, and the transformed joint density
function is a multivariate normal with mean $\tilde\mu$ and precision
matrix $Q$, i.e.
%
\begin{equation}
\log\tilde\pi(\tilde x | \theta,y) = \frac{1}{2}\log|Q(\theta)|-\frac
{1}{2}(\tilde x-\tilde\mu(\theta))'Q(\theta)(\tilde x-\tilde\mu(\theta
)) + \text{constant}
\label{eq:joint1}
\end{equation}
In the Laplace approximation defined in Equation~\eqref{eq:laplace},
both the numerator and the denominator should be evaluated in the point
$\tilde x = \mu(\theta)$, where $\mu(\theta)=(\mu_1(\theta),\ldots,\mu
_n(\theta))'$ is the mean of the Gaussian approximation $\tilde\pi
_G(x|\theta,y)$.
Thus, the density functions above should be evaluated in $\tilde x = \mu
(\theta)$.
From Equation~\eqref{eq:laplace}, the original (uncorrected) log
posterior is
%
\begin{equation}
\log\tilde\pi(\theta| y) = \log\pi({x},{\theta},{y}) - \log\tilde
\pi_G({x}|{\theta},{y}) + \text{constant}
\label{eq:joint2}
\end{equation}
evaluated at $x=\mu(\theta)$, where
%
\begin{equation}
\left.\log\tilde\pi_G({x}|{\theta},{y})\right\vert_{x= \mu(\theta)} =
\frac{1}{2}\log|Q(\theta)| + \text{constant},
\label{eq:joint3}
\end{equation}
Comparing equations~\eqref{eq:joint1}, \eqref{eq:joint2}, and \eqref
{eq:joint3}, we see that the copula approximation can be implemented by
adding the term $C(\theta)$ to the already calculated log posterior
evaluated at $\mu(\theta)$, where
\[
C(\theta)=\frac{1}{2}(\mu(\theta)-\tilde\mu(\theta))'Q(\theta)(\mu
(\theta)-\tilde\mu(\theta)).
\]
The addition of the term $C(\theta)$ does not add significantly to the
computational cost of INLA --- this simple operation is essentially free.

For the INLA implementation of the copula correction, we have found
that it is sufficient to only include fixed effects (including any
random effects of length one) in the calculation of $C(\theta)$.
The effect of the correction is strongest and most consistent for the
fixed effects, while the (often very numerous) random effects
contribute very small individual effects to the correction, mainly
adding extraneous noise to the estimation.
For these reasons, including only fixed effects gives better numerical
stability and also seems to provide a more accurate approximation,
while reducing computational costs.
Conceptually, including only the fixed effects involves finding $\Sigma
(\theta)=Q(\theta)^{-1}$, and then again finding
$Q_{\mathcal{J}}(\theta)=\Sigma_{\mathcal{J}}(\theta) ^{-1}$ (where
$\mathcal{J}$ is the index set of the fixed effects), which might seem
computationally costly. However, it can be done cheaply by using the
linear combination feature described in Section 4.4 of~\citet
{martins2013}: If $n_f$ is the number of fixed effects, only the
(parallel) solution of a $n_f$-dimensional linear system is needed.

Additionally, to guard against over-correction, we perform a soft
thresholding on $C(\theta)$, as follows: First we define a sigmoidal
function $f(t)$:
\[
f(t)=\frac{2}{1+\exp(-2t)} -1,
\]
which is increasing, has derivative equal to one at the point $t=0$,
and where $f(t) \rightarrow1$ as $t \rightarrow\infty$. Then we
replace $C(\theta)$ by $C_t(\theta)$, where
\[
C_t(\theta) = u f(C(\theta)/u),
\]
for $u = n_f \xi$ and with the ``correction factor'' parameter $\xi>0$
determining the degree of shrinkage (more shrinkage for smaller values
of $\xi$).
Since the function $f(t)$ is approximately linear with unit slope
around zero, $C_t(\theta)$ will be close to
$C(\theta)$ for small and moderate values of $C(\theta)$, while larger
values will be increasingly shrunk toward zero. Note that since
$f(t) < 1$ for all $t$,
$C_t(\theta) < u$.
The value of $\xi$ does not have a large impact on the results unless a
too small value is chosen. Its main purpose is as a safeguard to avoid
too large corrections in very difficult cases. In our experience $\xi
=1$ gives too strong shrinkage, while for example $\xi=100$ corresponds
to no shrinkage, so it seems clear that $\xi$ should be somewhere in
between these extremes.
We have found that $\xi=10$ is a good choice, letting the correction
do its job while guarding against too large changes, and we have used
this value for all of the examples. Results appear to be very robust to
the exact value chosen for $\xi$. Note also that since the correction
effect $u$ is scaled with the number $n_f$ of fixed effects, it is less
surprising that a single value for $\xi$ could work well in a wide
variety of circumstances.

As mentioned, we have also investigated a more general case of the
copula construction, where we retain $F_i$ as a skew normal
distribution, i.e.~the CDF of $\tilde\pi_{SN}(x_i | \theta,y)$. This
results in a more complicated correction term $C_{\text{skew}}(\theta
)$, derived in Appendix~\ref{app}. We have not found any appreciable
differences in the accuracy compared to the simpler case without
skewness, so we have concluded that the non-skew version is preferable
due to its simplicity.
We will show both the skew and the non-skew correction for the toenail
data discussed in Section~\ref{sec:toenail}, but otherwise we show only
results from the simpler non-skew version. We have tried both
corrections on many (both real and simulated) data sets, and never seen
a significant difference in the results.

\section{Empirical results}
\label{sec:examples}

\subsection{Have we solved the problems detected by Fong et al.~(2010)?}
\label{sec:simulation-study}

As mentioned in the Introduction, \citet{fong2010} studied the use of
INLA for binary valued GLMMs, and they showed that the approximations
were inaccurate in some cases. We have redone the simulation experiment
described on pages 10--14 of the Supplementary Material
of~\citet{fong2010}, both for INLA without any correction, and INLA
with the ``mean only'' correction described in Section~\ref{sec:method}.

In the original simulation study by \citet{fong2010}, $Y_{ij}$ are iid
$\mathrm{Binomial}(m,p_{ij})$, with $i=1,\ldots,100$ clusters,
$j=1,\ldots,7$ observations per cluster, and $m \in\{1,2,4,8\}$. Given
$x_i = 0$ for $i \leq50$ and $x_i = 1$ otherwise, and sampling
times $t = (t_1,\ldots,t_7)' = (-3, -2, -1, 0, 1, 2, 3)'$, the
following two models were considered:
%
\begin{eqnarray}
\label{eq:07}
\text{logit} \ p_{ij} &=& \beta_0 + \beta_1 t_j + \beta_2 x_i +
\beta_3 t_j x_i + b_{0i}\\
\label{eq:08}
\text{logit} \ p_{ij} &=& \beta_0 + \beta_1 t_j + \beta_2 x_i +
\beta_3 t_j x_i + b_{0i} + b_{1i}t_j,
\end{eqnarray}
which corresponds to models (0.7) and (0.8) on page 11 of the
Supplementary Material of~\citet{fong2010}.

We first consider model~\eqref{eq:07}.
We only show the results for $m = 1$ (i.e. binary data), as this is the
most difficult case with the largest errors in the
approximation. The correction also works well for $m > 1$, but this
case is easier to deal with for INLA. This is seen empirically, and is
also as expected based on considering the asymptotic properties of the
Laplace approximation: for $m > 1$ there is more ``borrowing of
strength''/replication, so the original approximation should work better.
We use the same settings as \citet{fong2010}: $b_{0i} \sim_{iid} N(0,
\sigma_0^2)$ where $\sigma_0^2 = 1$,
the prior $\text{Gamma}(0.5, 0.0164)$ for $\sigma_0^{-2}$ and
$N(0,1000)$ priors for the $\beta_i$. The true values of the fixed
effects are $\beta= (\beta_0, \beta_1, \beta_2, \beta_3)' = (-2.5,
1.0, -1.0, -0.5)'$. We made 1,000 simulated data sets, running INLA
both with and without the new correction, as well as very long MCMC
chains using JAGS (each of the 1,000 datasets were run with 1,000,000
MCMC samples after a burn-in of 100,000, using every 100th sample).

%
\begin{table}[t!]
\tabcolsep=0pt
\caption{Results from simulation study from model~\eqref{eq:07}}
\label{table1}
\begin{tabular*}{\textwidth}{@{\extracolsep{4in minus 4in}}|llrrrrrrrr|@{}}
\hline
\multicolumn{10}{|c|@{}}{\textbf{Averages of posterior means:}}\\
& & \multicolumn{1}{c}{$\sigma_0^2$}
& \multicolumn{1}{c}{$\sigma_0$}
& \multicolumn{1}{c}{$\log\sigma_0^{-2}$}
& \multicolumn{1}{c}{$\beta_0$}
& \multicolumn{1}{c}{$\beta_1$}
& \multicolumn{1}{c}{$\beta_2$}
& \multicolumn{1}{c}{$\beta_3$} & \multicolumn{1}{c|@{}}{}\\
& True values & 1.000 & 1.000 & 0.000 & -2.500 & 1.000 & -1.000 &
-0.500 & \\
& Uncorrected INLA & 0.705 & 0.722 & 1.133 & -2.494 & 0.998 & -1.052 &
-0.486 & \\
& Corrected INLA & 0.952 & 0.850 & 0.775 & -2.562 & 1.024 & -1.080 &
-0.504 & \\
& MCMC & 0.946 & 0.849 & 0.773 & -2.537 & 1.017 & -1.081 & -0.482 &\\
\multicolumn{10}{|c|@{}}{}\\
\multicolumn{10}{|c|@{}}{\textbf{Comparison between INLA and MCMC,
(E(INLA)-E(MCMC))/sd(MCMC):}}\\
& & \multicolumn{1}{c}{$\sigma_0^2$}
& \multicolumn{1}{c}{$\sigma_0$}
& \multicolumn{1}{c}{$\log\sigma_0^{-2}$}
& \multicolumn{1}{c}{$\beta_0$}
& \multicolumn{1}{c}{$\beta_1$}
& \multicolumn{1}{c}{$\beta_2$}
& \multicolumn{1}{c}{$\beta_3$} & \multicolumn{1}{c|@{}}{}\\
& Uncorrected INLA & -0.382 & -0.403 & 0.390 & 0.120 & -0.127 & 0.052 &
-0.016 &\\
& Corrected INLA & -0.003 & 0.000 & 0.002 & -0.073 & 0.046 & -0.002 &
-0.101 &\\
\multicolumn{10}{|c|@{}}{}\\
\multicolumn{10}{|c|@{}}{\textbf{Ratio of variances,
Var(INLA)/Var(MCMC):}}\\
& & \multicolumn{1}{c}{$\sigma_0^2$}
& \multicolumn{1}{c}{$\sigma_0$}
& \multicolumn{1}{c}{$\log\sigma_0^{-2}$}
& \multicolumn{1}{c}{$\beta_0$}
& \multicolumn{1}{c}{$\beta_1$}
& \multicolumn{1}{c}{$\beta_2$}
& \multicolumn{1}{c}{$\beta_3$} & \multicolumn{1}{c|@{}}{}\\
& Uncorrected INLA & 0.585 & 0.812 & 1.174 & 0.822 & 0.834 & 0.882 &
0.889 &\\
& Corrected INLA & 0.933 & 0.956 & 0.998 & 0.904 & 0.871 & 0.943 &
0.908 &\\
\multicolumn{10}{|c|@{}}{}\\
\multicolumn{10}{|c|@{}}{\textbf{Average coverage of 95\% intervals from
INLA over MCMC samples:}}\\
& & \multicolumn{1}{c}{$\sigma_0^2$}
& \multicolumn{1}{c}{$\sigma_0$}
& \multicolumn{1}{c}{$\log\sigma_0^{-2}$}
& \multicolumn{1}{c}{$\beta_0$}
& \multicolumn{1}{c}{$\beta_1$}
& \multicolumn{1}{c}{$\beta_2$}
& \multicolumn{1}{c}{$\beta_3$} & \multicolumn{1}{c|@{}}{}\\
& Uncorrected INLA & 90.3\% & 90.0\% & 90.8\% & 92.6\% & 92.7\% &
93.5\% & 93.5\%& \\
& Corrected INLA & 94.2\% & 93.9\% & 94.4\% & 93.5\% & 93.1\% & 94.3\%
& 93.7\% &\\
\hline
\end{tabular*}
\end{table}

Results from the simulation study are shown in Table~\ref{table1}. Note
that the aim here is to be as close as possible to the MCMC results,
not the true values. The upper part of the table shows averages of the
posterior means over the 1,000 simulations. We see that INLA gets much
closer to the MCMC results for all parameters except $\beta_3$, which
is in any case reasonably close to the MCMC value. The improvement is
particularly large for the variance parameter. This is also seen in the
second panel, which shows $(\text{E}(\theta^{\text{INLA}}|y)-\text
{E}(\theta^{\text{MCMC}}|y))/\text{sd}(\theta^{\text{MCMC}}|y)$ for
each parameter $\theta$ (averaged over the 1,000 simulations), i.e.~the
difference in INLA and MCMC estimates scaled by the MCMC standard
deviation. Here, the random effects variance and the fixed effects
except $\beta_3$ are also more accurately estimated. The third lower
panel shows the ratios
$\text{Var}(\theta^{\text{INLA}}|y)/\text{Var}(\theta^{\text
{MCMC}}|y)$; here the correction improves the estimation of the
variance for all parameters. For $\sigma_0^2$, $\sigma_0$ and $\log
\sigma_0^{-2}$ we get very close to a ratio of one, and there are also
major improvements for the fixed effects variances.
Finally, the bottom panel shows average coverage of 95\% (i.e.,
$(q_{0.025},q_{0.975})$) credible intervals from INLA over the MCMC
samples for each simulated data set. Clearly, coverage is improved
considerably by the correction.
Table~\ref{table2} reports summary statistics for the computation times
in seconds over the 1,000 data sets. Note that in this case
computational times are abnormally high due to somewhat extreme
parameter settings -- INLA will usually be much faster. However, ratios
of computing times for the corrected vs the uncorrected versions should
stay approximately the same.

Appendix~\ref{appSim} contains additional simulation studies:
Results from simulations for model~\eqref{eq:08} for different values
of the covariance matrix of $(b_{0i},b_{1i})'$ are shown in
Appendix~\ref{sec:model-with-two}. Furthermore, in Appendix~\ref
{sec:model-with-very} we consider the effect of having extremely few
observations per cluster, while we in Appendix~\ref
{sec:simul-with-missp} study a misspecified model, simulating from
model~\eqref{eq:08} while estimating model~\eqref{eq:07}. The
correction appears to work well for all the different cases considered
in Appendix~\ref{appSim}.

%
\begin{table}[t]
\caption{Summary statistics for computation times in seconds for each
data set}
\label{table2}
\begin{tabular}{|lcccccc|@{}}
\hline
& Min. & 1st Qu. & Median & Mean & 3rd Qu. & Max. \\
\hline
Uncorrected INLA & 1.70 & 2.25 & 2.39 & 2.43 & 2.56 & 3.41 \\
Corrected INLA & 1.77 & 2.58 & 2.69 & 2.74 & 2.86 & 3.89 \\
\hline
\end{tabular}
\end{table}

%
\begin{figure}[t!]
\includegraphics[width=\textwidth]{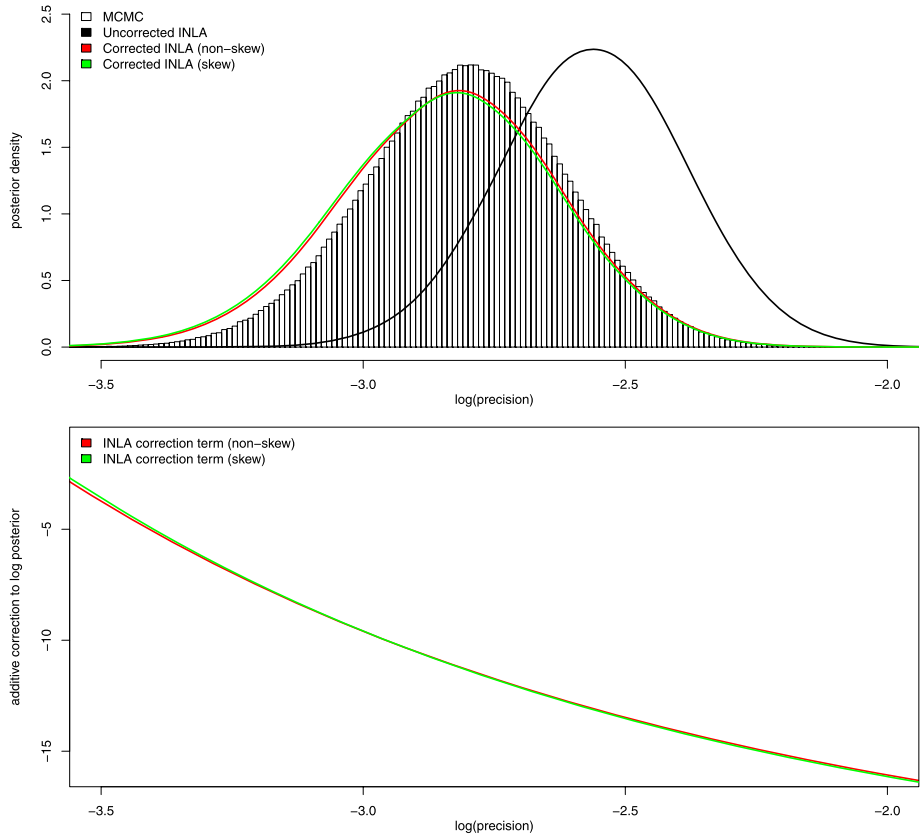}
\caption{The top panel shows the posterior density of the
hyperparameter (log precision) for the toenail data. The histogram is
from one million MCMC samples from JAGS, the blue curve is from
uncorrected INLA, the red curve from INLA with the ``mean only''
correction, and the green curve from INLA with the ``mean and skewness''
correction. The bottom panel shows the additive corrections to the log
posterior for the toenail data for the ``mean only'' and the ``mean and
skewness'' correction.}
\label{figure3}
\end{figure}

\vspace*{-3pt}
\subsection{What happens when the random effects variance increases?}
\label{sec:toenail}
\vspace*{-3pt}

We start by discussing the toenail data, which is a classical data set
with a binary response and repeated measures~\citep
{toenaildata,toenailpaper}. The data are from a clinical trial
comparing two competing treatments for toenail infection (dermatophyte
onychomycosis). The 294 patients were randomized to receive either
itraconazole or terbinafine, and the outcome (either ``not severe''
infection, coded as $0$, or ``severe'' infection, coded as $1$) was
recorded at seven follow-up visits. Not all patients attended all the
follow-up visits, and the patients did not always appear at the
scheduled time. The exact time of the visits (in months since baseline)
was recorded.
For individual $i$, visit $j$, with outcome $y_{ij}$, treatment $\text
{Trt}_i$ and time $\text{Time}_{ij}$ our model is then\begingroup\abovedisplayskip=0pt\belowdisplayskip=0pt
\begin{eqnarray*}
y_{ij} &\sim& \text{Bernoulli}(p_{ij})\\[-1pt]
\text{logit} \ p_{ij} &=& \alpha_0 + \alpha_{\text{Trt}} \text{Trt}_i +
\alpha_{\text{Time}} \text{Time}_{ij} + \alpha_{TT} \text{Trt}_i \text
{Time}_{ij} + b_i\\[-1pt]
b_i &\sim& N(0, \sigma^2).
\end{eqnarray*}\endgroup
Notice that this is the same model as model~\eqref{eq:07}, except that
the time variable here varies over individuals.
Normal priors with mean zero and variance $10^4$ were used for $\alpha
_0$, $\alpha_{\text{Trt}}$, $\alpha_{\text{Time}}$, and $\alpha_{TT}$.
INLA underestimates $\sigma^2$ quite severely. The top panel of
Figure~\ref{figure3} shows the different estimates of the posterior
distribution of the log precision, $\log \sigma^{-2}$. The histogram
shows the results from a long MCMC run using JAGS, the black curve
shows the posterior from INLA without the correction, the red curve
shows the simple (non-skew) version of the INLA correction, while the
green curve shows the INLA correction accounting for skewness (as
discussed in Appendix \ref{app}).
The bottom panel of Figure~\ref{figure3} shows the additive correction
to the log posterior density, as a function of the hyperparameter (log
precision). We see that there is very little difference between the two
corrections.

\begin{figure}[t!]
\includegraphics[width=\textwidth]{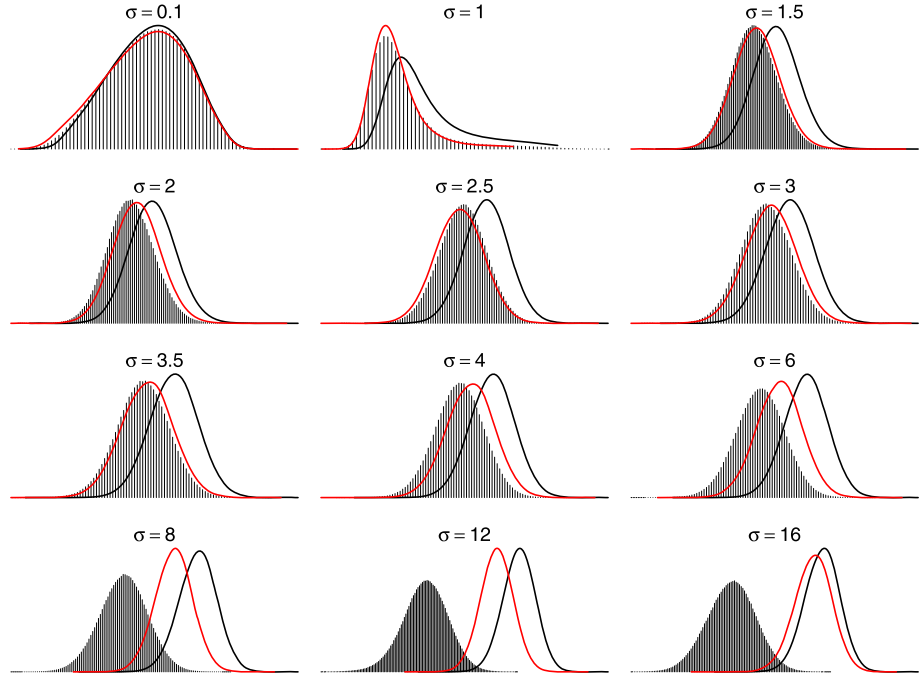}
\caption{Posteriors densities of log precision from the toenail
simulation experiment for different values of the random effects
standard deviation $\sigma$ (the value of $\sigma$ is shown above each
histogram). The histograms are from long MCMC runs, uncorrected INLA
are shown as black curves, while the red curves shows INLA with the
correction. Since the goal here is to study the difference between MCMC
and INLA, we omit axes -- the relevant scale is given by the MCMC
variances, which are evident from the histograms.}
\label{figure4}
\end{figure}

For the toenail data, the estimated random effects standard deviation
is approximately $\sigma=4$, which is very high. To investigate how the
copula correction works as $\sigma$ increases, we studied simulated
data sets from the model above, where we set $\sigma$ to different
values, and where the $\alpha$ parameters were fixed to the values from
a long MCMC run using the real toenail data (i.e., we simulate only the
outcome, keeping the covariates unchanged). Results are shown in
Figure~\ref{figure4} for different values of $\sigma$ ranging from
$\sigma=1$ to $\sigma=16$. We clearly get very accurate corrected
posteriors for $\sigma<4$. For $\sigma\geq4$, we gradually get a
tendency of under-correction.

%

\subsection{Simulated example with Poisson likelihood}
\label{sec:poisson}

\begin{figure}[t!]
\includegraphics[width=\textwidth]{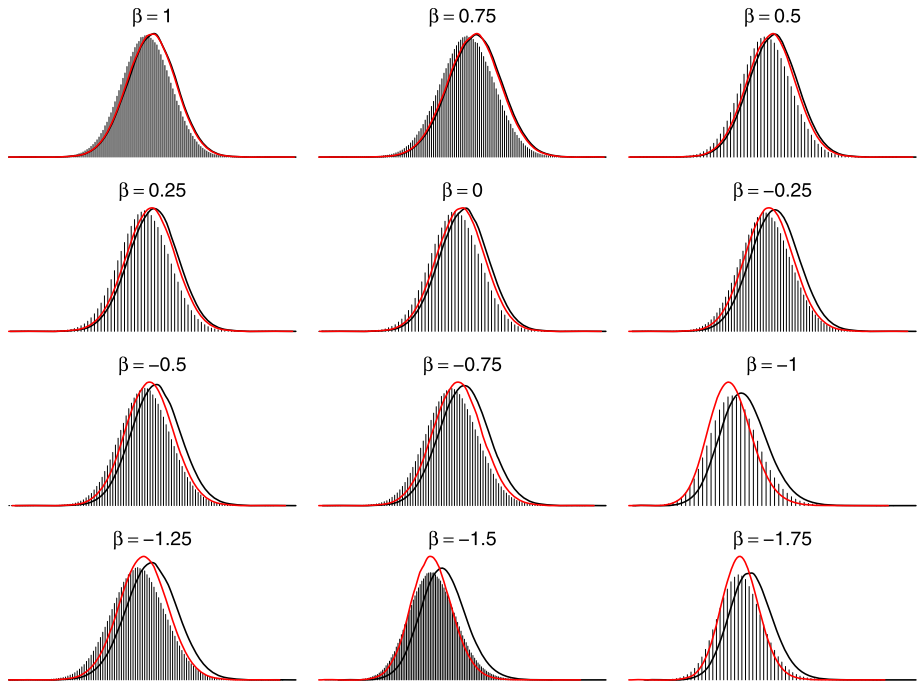}
\caption{Posteriors densities of log precision from the Poisson
simulation experiment for different values of the intercept $\beta$
(the value of $\beta$ is shown above each histogram). The histograms
are from long MCMC runs, uncorrected INLA posteriors are shown as black
curves, while the red curves shows INLA posteriors with the correction.}
\label{figure5}
\end{figure}

We shall now study the case where the data are Poisson distributed.
We consider a simple simulated (extreme) example in order to
investigate how well the correction works in the Poisson case.
For $i=1,\ldots,n$ we generated iid $y_i \sim\text{Poisson}(\mu_i)$ where
\[
\log(\mu_i) = \beta+ u_i,
\]
with $u_i \sim N(0,\sigma^2)$. We chose $n=300$ and $\sigma^2=1$, a
$\text{Gamma}(1,1)$ prior for the precision $\sigma^{-1}$, and a
$N(0,1)$ prior for $\beta$.
Figure~\ref{figure5} shows the results for different values of the
intercept $\beta$. Each histogram is based on ten parallell MCMC runs
using JAGS, each with 200,000 iterations after a burn-in of 100,000.
Here, reduction of $\beta$ implies that estimation is more difficult,
since negative $\beta$ with large absolute value will imply that the
counts $y_i$ are very low, with many zeroes, and the data are
uninformative. We see that uncorrected INLA tends to get less accurate
as $\beta$ moves towards more extreme values, while the correction
seems to work well.

%

\subsection{What if the latent structures are more complicated?}

Until now, we have considered fairly simple latent structures, where
the random effects have been iid (multivariate) normally distributed.
The reader may perhaps wonder if the generality implied by having
``latent Gaussian models'' in the title is really justified -- what if
latent structures are more complicated, with for example temporal or
spatial structure?\vfill\eject

In fact, the complexity of the latent field is not particularly
relevant for the accuracy problems we study here. This can be seen by
considering the basic formulation of the latent Gaussian model together
with the main building blocks of the INLA machinery: Essentially, the
latent structure is contained within the Gaussian part, for which the
computations are exact (and fast, since the precision matrix of the
Gaussian part will usually be sparse). In a sense, the LGM approach
separates the estimation in an ``easy'' (Gaussian) part and a
``difficult'' part. It is perhaps somewhat counterintuitive at first
sight that the dynamic/time-series/spatial model constitutes the
``easy'' part! In this paper we have in fact considered the
``difficult'' part, aiming to choose examples at the boundary of what
we considered to be feasible. Thus, we argue that our general title is
indeed justified.

\begin{figure}[p!]
\includegraphics[width=\textwidth]{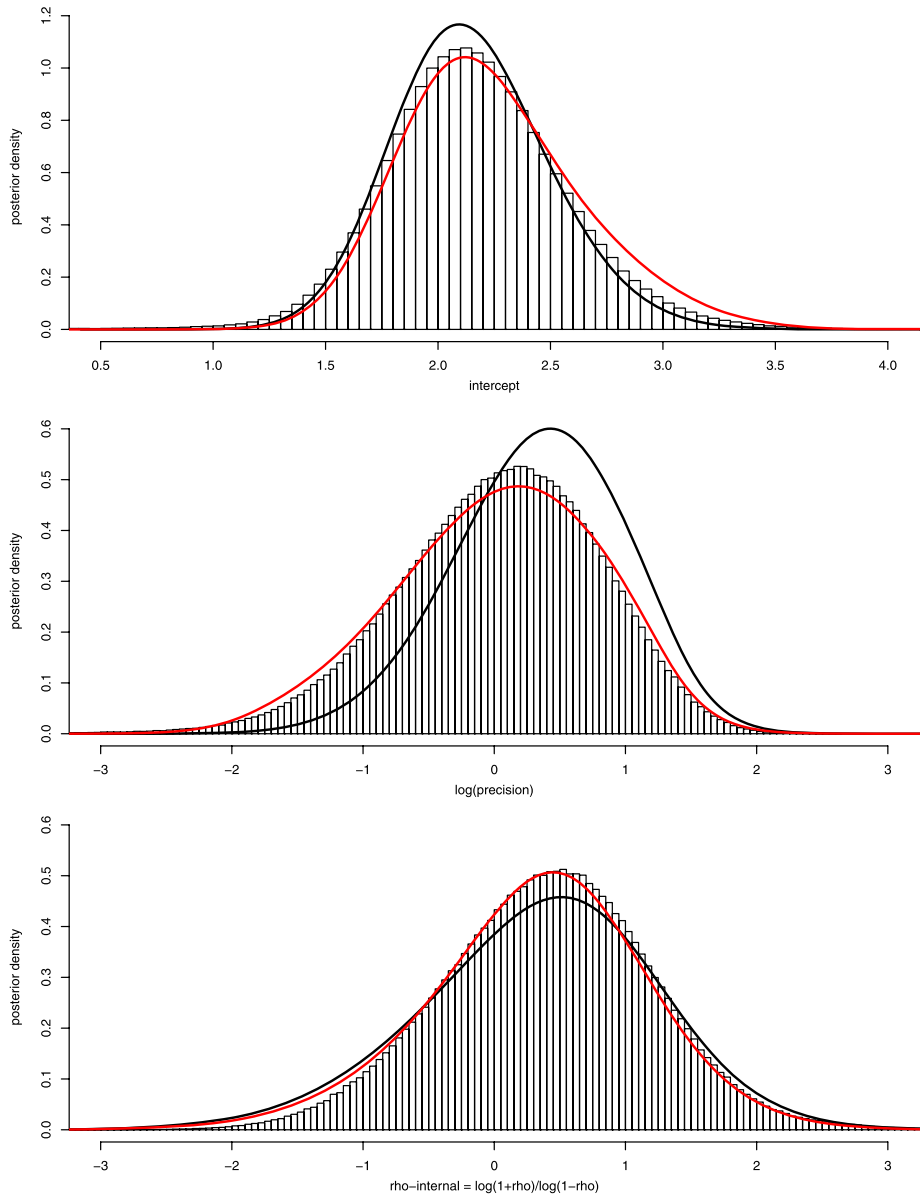}
\caption{AR1 example defined in Equation~\eqref{eq:ar1}. The top panel
displays results for the intercept
parameter $\beta$, the middle panel shows results for $\theta_1 =
log(\kappa)$, and the bottom panel show results
for $\theta_2=\log\left(\frac{1+\rho}{1-\rho}\right)$, the ``internal
$\rho$'' of INLA. The histograms show posterior distributions from a
long MCMC run (ten chains of one million iterations each), the black
curves show the posterior from INLA, while the red curve shows the
posteriors using our new correction to INLA.}
\label{figure6}
\end{figure}

We illustrate this with a simple simulated example where the latent
structure is auto-regressive of order one (AR1), using a similar setup
as in the ``minimal'' example presented in the Introduction.
For $i=1,2,\ldots,n$, let $\mathrm{Prob}(y_i=0)=1-p_i$, $\mathrm
{Prob}(y_i=1)=p_i$, and
%
\begin{equation}
\label{eq:ar1}
\mathrm{logit}(p_i) = \beta+ u_i,
\end{equation}
where the $u_i$ are now given an AR1 model, as follows: $u_1 \sim
N(0,\kappa^{-1})$, $u_i = \rho u_{i-1}+ \varepsilon_i$ (where
$\varepsilon_i \sim N(0,\tau^{-1})$) for $i = 2,3,\ldots,n$, where
$\kappa= \tau(1-\rho^2)$ is the marginal precision. Define
\[
\theta_1=\log(\kappa)
\]
and
\[
\theta_2=\log\left(\frac{1+\rho}{1-\rho}\right)
\]
which is the parameterization used internally in INLA. We use a
Gamma$(1, 1)$ prior for $\kappa$, and $N(0,1)$ priors for both $\beta$
and $\theta_2$. Data was simulated from model~\eqref{eq:ar1}, using
$\rho= 0.5$, $\tau= 1$ and $\beta= 2$ as the true values. As in the
example in the Introduction, we made long MCMC runs and compared the
results to INLA both with and without the correction. The results are
shown in Figure~\ref{figure6}. Again, it is clear that the overall
accuracy of INLA is improved using the correction.

%

\section{Discussion}
\label{sec:conclusion}

The binary (and, more generally, binomial) GLMMs discussed in
Sections~\ref{sec:simulation-study} and~\ref{sec:toenail} are are
important in many applications, particularly for biomedical data.
Poisson GLMMs are also of great interest, and among the difficult cases
here are point processes such as log-Gaussian Cox processes~\citep
{illian2012,simpson2013}, where data are typically extremely sparse:
essentially there are ones at the observed points, and zeroes
everywhere else. Our example in Section~\ref{sec:poisson} shows a
stylized, extreme case of this type. Studying the correction for the
full log-Gaussian Cox process case could be a topic for future work.
Even though the point process case may be difficult, there will often
be some degree of smoothing and/or replication making inference easier,
so real data sets should be less extreme than the simulated example in
Section~\ref{sec:poisson}.

From the results in this paper, it appears that the copula correction
is robust and works well. There is no general theory guaranteeing that
the method will always work under all circumstances, but we feel that
the intuition underlying the method is quite strong.
Since INLA for LGMs is quite accurate in most cases, the correction is
not needed in general, only for problematic cases such as those
discussed in this paper. Using the copula correction method, we can
stretch the limits of applicability of INLA, while maintaining its
computational speed.


%
\begin{appendix}

\section{Copula correction accounting for skewness}\label{app}

Let $\tilde F_i$ denote the ``standardized'' skew normal CDF
corresponding to\break $\tilde\pi_{SN}(x_i | \theta,y)$.
We start by finding the Jacobian of the transformation defined in
Equation~\eqref{eq:copula}.
Note first that, immediately from~\eqref{eq:copula}
\[
\tilde F_i\left(\frac{\tilde x_i-\tilde\mu_i(\theta)}{\sigma_i(\theta
)}\right) = \Phi\left(\frac{x_i-\mu_i(\theta)}{\sigma_i(\theta)}\right).
\]
Letting $\tilde f_i$ and $\phi$ denote the density functions
corresponding to $\tilde F_i$ and $\Phi$,
differentiating with respect to $\tilde x_i$ then gives
\[
\tilde f_i\left(\frac{\tilde x_i-\tilde\mu_i(\theta)}{\sigma_i(\theta
)}\right) \frac{1}{\sigma_i(\theta)}
= \phi\left(\frac{x_i-\mu_i(\theta)}{\sigma_i(\theta)}\right) \frac
{\partial x_i}{\partial\tilde x_i}\frac{1}{\sigma_i(\theta)}
\]
so
\[
\frac{\partial x_i}{\partial\tilde x_i}=
\frac{\tilde f_i\left(\frac{\tilde x_i-\tilde\mu_i(\theta)}{\sigma
_i(\theta)}\right)}{\phi\left(\frac{x_i-\mu_i(\theta)}{\sigma_i(\theta
)}\right)}
= \frac{\tilde f_i\left(\frac{\tilde x_i-\tilde\mu_i(\theta)}{\sigma
_i(\theta)}\right)}
{\phi\left(\Phi^{-1} \left[\tilde F_i \left(\frac{\tilde x_i-\tilde\mu
_i(\theta)}{\sigma_i(\theta)}\right)\right]\right)}
\]
and the Jacobian of the transformation is
$\prod_{i=1}^n \frac{\partial x_i}{\partial\tilde x_i}$
since $\frac{\partial x_i}{\partial\tilde x_j}=0$ for $i \neq j$ and
$\frac{\partial x_i}{\partial\tilde x_i}\geq0$ for all $i$.

Note that $(x-\mu(\theta))'Q(\theta)(x-\mu(\theta))=\sum_{i=1}^n \sum
_{j=1}^n Q_{ij}(\theta) (x_i-\mu_i(\theta))(x_j-\mu_j(\theta))$ where
$Q(\theta)=(Q_{ij}(\theta))$.
Collecting the different terms and again substituting
\[
\frac{x_i-\mu_i(\theta)}{\sigma_i(\theta)}=\Phi^{-1} \left[\tilde F_i
\left(\frac{\tilde x_i-\tilde\mu_i(\theta)}{\sigma_i(\theta)}\right
)\right],
\]
the transformed joint log density $\tilde\pi(\tilde x | \theta,y)$ is therefore
\begin{gather*}
\log\tilde\pi(\tilde x | \theta,y) = \frac{1}{2}\log|Q(\theta)|\\
-\frac{1}{2}\sum_{i=1}^n \sum_{j=1}^n Q_{ij}(\theta) \sigma_i(\theta)
\sigma_j
\Phi^{-1} \left[\tilde F_i \left(\frac{\tilde x_i-\tilde\mu_i(\theta
)}{\sigma_i(\theta)}\right)\right]
\Phi^{-1} \left[\tilde F_j \left(\frac{\tilde x_j-\tilde\mu_j(\theta
)}{\sigma_j(\theta)}\right)\right]\\
+ \sum_{i=1}^n \log\tilde f_i\left(\frac{\tilde x_i-\tilde\mu_i(\theta
)}{\sigma_i(\theta)}\right)
- \sum_{i=1}^n \log\phi\left(\Phi^{-1} \left[\tilde F_i \left(\frac
{\tilde x_i-\tilde\mu_i(\theta)}{\sigma_i(\theta)}\right)\right]\right)
+ \text{constant}
\end{gather*}

The original (uncorrected) log posterior is
\[
\log\tilde\pi(\theta| y) = \log\pi({x},{\theta},{y}) - \log\tilde
\pi_F({x}|{\theta},{y}) + \text{constant},
\]
evaluated at $x=\mu(\theta)$ where $ \left.\log\tilde\pi_F({x}|{\theta
},{y})\right\vert_{x= \mu(\theta)} = \frac{1}{2}\log|Q(\theta)| + \text
{constant}$. Therefore, the version of the copula correction accounting
for skewness amounts to adding a term $C_{\text{skew}}(\theta)$ to the
original log joint posterior, where
\begin{gather*}
C_{\text{skew}}(\theta) =\\
\frac{1}{2}\sum_{i=1}^n \sum_{j=1}^n Q_{ij}(\theta) \sigma_i(\theta)
\sigma_j(\theta)
\Phi^{-1} \left[\tilde F_i \left(\frac{\mu_i(\theta)-\tilde\mu_i(\theta
)}{\sigma_i(\theta)}\right)\right]\!
\Phi^{-1} \!\left[\tilde F_j \left(\frac{\mu_j(\theta)-\tilde\mu_j(\theta
)}{\sigma_j(\theta)}\right)\right]\\
+ \sum_{i=1}^n \log\tilde f_i\left(\frac{\mu_i(\theta)-\tilde\mu
_i(\theta)}{\sigma_i(\theta)}\right)
- \sum_{i=1}^n \log\phi\left(\Phi^{-1} \left[\tilde F_i \left(\frac{\mu
_i(\theta)-\tilde\mu_i(\theta)}{\sigma_i(\theta)}\right)\right]\right).
\end{gather*}

Calculations of $\tilde F_i$ and $\tilde f_i$ were done using the
functions \texttt{psn} and \texttt{dsn} in the R package \texttt
{sn}~\citep{sn}.

\section{Additional simulation results}\label{appSim}

\subsection{Model with two dependent random effects}
\label{sec:model-with-two}

Here we study model (0.8) from page 11 of the Supplementary Material
of~\citet{fong2010}, where the observations $Y_{ij}$ are iid
Binomial$(p_{ij}, m)$, with $i = 1, \ldots, 100$ clusters, $j =
1,\ldots,7$ observations per cluster, $x_i = 0$ for $i \leq50$ and
$x_i = 1$ otherwise, and sampling times $t = (t_1,\ldots,t_7)' =
(-3,-2,-1,0,1,2,3)'$. The model is
%
\begin{equation}
\label{eq:fongbivar}
\text{logit} \ p_{ij} = \beta_0 + \beta_1 t_j + \beta_2 x_i +
\beta_3 t_j x_i + b_{0i} + b_{1i}t_j,
\end{equation}
where the $(b_{0i},b_{1i})'$ are iid bivariate normally distributed
with mean $(0,0)$. Following~\citet{fong2010}, the prior for the
precision matrix of $(b_{0i},b_{1i})'$ is a Wishart distribution with
three degrees of freedom and diagonal scale parameter with diagonal
elements $0.17$ and $0.025$. The fixed effects are given $N(0,1000)$ priors.

As in~\citet{fong2010}, we shall consider the case when $b_{0i}$ and
$b_{1i}$ are uncorrelated, but we shall here also consider the
correlated case with correlation $\rho= 0.5$ and $\rho= 0.9$,
respectively. Additionally, we consider two different settings of the
marginal variances of $b_{0i}$ and $b_{1i}$:
\begin{enumerate}
\item Var$(b_{0i}) = 0.5$, Var$(b_{1i}) = 0.25$ (as in~\citet{fong2010}),
\item Var$(b_{0i}) = 3.0$, Var$(b_{1i}) = 0.5$.
\end{enumerate}

For each of the two settings of the marginal variances above, we ran
the simulation experiment for the three settings of $\rho$ ($0$, $0.5$
and $0.9$ correlation), giving six simulation settings in total. For
each simulation setting, we made 200 simulated data sets, and ran two
MCMC chains of 200,000 iterations each (after discarding the first
100,000 iterations) for each simulated data set.

We have yet to specify the number of trials $m$ in the binomial
distribution. It turns out that this model is nearly unidentifiable for
$m=1$, with very slow MCMC convergence and with numerical instability
when running INLA (both with and without the correction). Therefore, we
will here consider $m>1$, and show the results for $m=2$. Results (not
shown) are similar also for $m>2$. As expected, the estimation becomes
more accurate as $m$ grows, and for large $m$ (say, $m \geq10$) there
is no need for the INLA correction anymore.

Results are shown in Tables~\ref{table3}--\ref{table8} below, where we
use the parameterization $(\theta_1,\theta_2,\theta_3)$ used internally
by INLA, where
$\theta_1=\log(\sigma_0^{-2})$, $\theta_2=\log(\sigma_1^{-2})$ and
$\theta_3=\log\left(\frac{1+\rho}{1-\rho}\right)$ (note that
$\theta_1$, $\theta_2$ and $\theta_3$ are defined on the whole real line).
It seems like the correction is working quite well, giving an overall
improvement. The coverage probabilities are improved in all cases
expect for the $\log\left(\frac{1+\rho}{1-\rho}\right)$ with $\rho\leq
0.5$, so we see an improvement for $38$ of the $6*7=42$ combinations of
parameters and simulation settings. The variance ratio
Var(INLA)/Var(MCMC) is also improved for nearly all the cases, while
the other performance measures show an overall (though not uniform)
improvement. The method does not seem to deteriorate for higher values
of the marginal variances and correlation.

%
\begin{table}[h!]
\vspace*{18pt}
\tabcolsep=4pt
\caption{Results from simulation study~\eqref{eq:fongbivar} with
Var$(b_{0i}) = 0.5$, Var$(b_{1i}) = 0.25$ and $\rho=0$}
\label{table3}
\renewcommand*{\arraystretch}{1.1}
\begin{tabular*}{\textwidth}{@{\extracolsep{\fill}}|llcccccccc|@{}}
\hline
\multicolumn{10}{|c|@{}}{\textbf{Averages of posterior means:}}\\
& & \multicolumn{1}{c}{$\log(\sigma_0^{-2})$}
& \multicolumn{1}{c}{$\log(\sigma_1^{-2})$}
& \multicolumn{1}{c}{$\log\left(\frac{1+\rho}{1-\rho}\right)$}
& \multicolumn{1}{c}{$\beta_0$}
& \multicolumn{1}{c}{$\beta_1$}
& \multicolumn{1}{c}{$\beta_2$}
& \multicolumn{1}{c}{$\beta_3$} & \multicolumn{1}{c|@{}}{}\\
& True values & 0.693 & 1.386 & 0.000 & -2.500 & 1.000 & -1.000 &
-0.500 & \\
& Uncorrected INLA & 1.527 & 2.130 & 1.566 & -2.664 & 1.021 & -0.692 &
-0.428 &\\
& Corrected INLA & 1.380 & 2.016 & 1.313 & -2.703 & 1.038 & -0.704 &
-0.429 &\\
& MCMC & 1.449 & 2.022 & 1.707 & -2.694 & 1.032 & -0.707 & -0.421 &\\

\multicolumn{10}{|c|@{}}{}\\
\multicolumn{10}{|c|@{}}{\textbf{Comparison between INLA and MCMC,
(E(INLA)-E(MCMC))/sd(MCMC):}}\\
& & \multicolumn{1}{c}{$\log(\sigma_0^{-2})$}
& \multicolumn{1}{c}{$\log(\sigma_1^{-2})$}
& \multicolumn{1}{c}{$\log\left(\frac{1+\rho}{1-\rho}\right)$}
& \multicolumn{1}{c}{$\beta_0$}
& \multicolumn{1}{c}{$\beta_1$}
& \multicolumn{1}{c}{$\beta_2$}
& \multicolumn{1}{c}{$\beta_3$} & \multicolumn{1}{c|@{}}{}\\
&Uncorrected INLA & 0.113 & 0.167 & -0.124 & 0.125 & -0.086 & 0.042 &
-0.038 &\\
& Corrected INLA & -0.086 & -0.028 & -0.329 & -0.028 & 0.045 & 0.007 &
-0.047 &\\
\multicolumn{10}{|c|@{}}{}\\
\multicolumn{10}{|c|@{}}{\textbf{Ratio of variances,
Var(INLA)/Var(MCMC):}}\\
& & \multicolumn{1}{c}{$\log(\sigma_0^{-2})$}
& \multicolumn{1}{c}{$\log(\sigma_1^{-2})$}
& \multicolumn{1}{c}{$\log\left(\frac{1+\rho}{1-\rho}\right)$}
& \multicolumn{1}{c}{$\beta_0$}
& \multicolumn{1}{c}{$\beta_1$}
& \multicolumn{1}{c}{$\beta_2$}
& \multicolumn{1}{c}{$\beta_3$} & \multicolumn{1}{c|@{}}{}\\
& Uncorrected INLA & 0.882 & 0.986 & 0.927 & 0.905 & 0.907 & 0.937 &
0.948 &\\
& Corrected INLA & 0.943 & 0.974 & 0.996 & 0.994 & 0.982 & 0.977 &
0.987 &\\

\multicolumn{10}{|c|@{}}{}\\
\multicolumn{10}{|c|@{}}{\textbf{Average coverage of 95\% intervals from
INLA over MCMC samples:}}\\
& & \multicolumn{1}{c}{$\log(\sigma_0^{-2})$}
& \multicolumn{1}{c}{$\log(\sigma_1^{-2})$}
& \multicolumn{1}{c}{$\log\left(\frac{1+\rho}{1-\rho}\right)$}
& \multicolumn{1}{c}{$\beta_0$}
& \multicolumn{1}{c}{$\beta_1$}
& \multicolumn{1}{c}{$\beta_2$}
& \multicolumn{1}{c}{$\beta_3$} & \multicolumn{1}{c|@{}}{}\\
& Uncorrected INLA & 92.4\% & 93.3\% & 92.8\% & 93.7\% & 93.7\% &
94.2\% & 94.3\% &\\
& Corrected INLA & 92.9\% & 93.7\% & 90.0\% & 93.7\% & 93.8\% & 94.6\%
& 94.7\% &\\
\hline
\end{tabular*}
\end{table}
\vfill

%
\begin{table}[h!]
\tabcolsep=4pt
\caption{Results from simulation study~\eqref{eq:fongbivar} with
Var$(b_{0i}) = 0.5$, Var$(b_{1i}) = 0.25$ and $\rho=0.5$}
\label{table4}
\renewcommand*{\arraystretch}{1.1}
\begin{tabular*}{\textwidth}{@{\extracolsep{\fill}}|llcccccccc|@{}}
\hline
\multicolumn{10}{|c|@{}}{\textbf{Averages of posterior means:}}\\
& & \multicolumn{1}{c}{$\log(\sigma_0^{-2})$}
& \multicolumn{1}{c}{$\log(\sigma_1^{-2})$}
& \multicolumn{1}{c}{$\log\left(\frac{1+\rho}{1-\rho}\right)$}
& \multicolumn{1}{c}{$\beta_0$}
& \multicolumn{1}{c}{$\beta_1$}
& \multicolumn{1}{c}{$\beta_2$}
& \multicolumn{1}{c}{$\beta_3$} & \multicolumn{1}{c|@{}}{}\\
&True values & 0.693 & 1.386 & 1.099 & -2.500 & 1.000 & -1.000 & -0.500
&\\
& Uncorrected INLA & 1.444 & 1.866 & 2.051 & -2.623 & 1.139 & -0.721 &
-0.590 &\\
& Corrected INLA & 1.366 & 1.786 & 1.900 & -2.642 & 1.148 & -0.730 &
-0.594 &\\
& MCMC & 1.363 & 1.790 & 2.176 & -2.649 & 1.148 & -0.731 & -0.582 &\\
\multicolumn{10}{|c|@{}}{}\\
\multicolumn{10}{|c|@{}}{\textbf{Comparison between INLA and MCMC,
(E(INLA)-E(MCMC))/sd(MCMC):}}\\
& & \multicolumn{1}{c}{$\log(\sigma_0^{-2})$}
& \multicolumn{1}{c}{$\log(\sigma_1^{-2})$}
& \multicolumn{1}{c}{$\log\left(\frac{1+\rho}{1-\rho}\right)$}
& \multicolumn{1}{c}{$\beta_0$}
& \multicolumn{1}{c}{$\beta_1$}
& \multicolumn{1}{c}{$\beta_2$}
& \multicolumn{1}{c}{$\beta_3$} & \multicolumn{1}{c|@{}}{}\\
& Uncorrected INLA & 0.126 & 0.134 & -0.118 & 0.111 & -0.073 & 0.027 &
-0.041& \\
& Corrected INLA & 0.013 & -0.031 & -0.252 & 0.035 & -0.005 & 0.002 &
-0.066 &\\
\multicolumn{10}{|c|@{}}{}\\
\multicolumn{10}{|c|@{}}{\textbf{Ratio of variances,
Var(INLA)/Var(MCMC):}}\\
& & \multicolumn{1}{c}{$\log(\sigma_0^{-2})$}
& \multicolumn{1}{c}{$\log(\sigma_1^{-2})$}
& \multicolumn{1}{c}{$\log\left(\frac{1+\rho}{1-\rho}\right)$}
& \multicolumn{1}{c}{$\beta_0$}
& \multicolumn{1}{c}{$\beta_1$}
& \multicolumn{1}{c}{$\beta_2$}
& \multicolumn{1}{c}{$\beta_3$} & \multicolumn{1}{c|@{}}{}\\
&Uncorrected INLA & 0.891 & 0.998 & 0.930 & 0.904 & 0.912 & 0.934 &
0.953& \\
& Corrected INLA & 0.948 & 1.001 & 1.043 & 0.942 & 0.953 & 0.952 &
0.979 &\\
\multicolumn{10}{|c|@{}}{}\\
\multicolumn{10}{|c|@{}}{\textbf{Average coverage of 95\% intervals from
INLA over MCMC samples:}}\\
& & \multicolumn{1}{c}{$\log(\sigma_0^{-2})$}
& \multicolumn{1}{c}{$\log(\sigma_1^{-2})$}
& \multicolumn{1}{c}{$\log\left(\frac{1+\rho}{1-\rho}\right)$}
& \multicolumn{1}{c}{$\beta_0$}
& \multicolumn{1}{c}{$\beta_1$}
& \multicolumn{1}{c}{$\beta_2$}
& \multicolumn{1}{c}{$\beta_3$} & \multicolumn{1}{c|@{}}{}\\
& Uncorrected INLA & 92.8\% & 94.1\% & 93.1\% & 93.8\% & 93.9\% &
94.2\% & 94.4\% & \\
& Corrected INLA & 93.7\% & 94.8\% & 92.8\% & 93.9\% & 94.1\% & 94.4\%
& 94.6\% &\\
\hline
\end{tabular*}
\end{table}

%
\begin{table}[t!]
\tabcolsep=4pt
\caption{Results from simulation study~\eqref{eq:fongbivar} with
Var$(b_{0i}) = 0.5$, Var$(b_{1i}) = 0.25$ and $\rho=0.9$}
\label{table5}
\renewcommand*{\arraystretch}{1.1}
\begin{tabular*}{\textwidth}{@{\extracolsep{\fill}}|llcccccccc|@{}}
\hline
\multicolumn{10}{|c|@{}}{\textbf{Averages of posterior means:}}\\
& & \multicolumn{1}{c}{$\log(\sigma_0^{-2})$}
& \multicolumn{1}{c}{$\log(\sigma_1^{-2})$}
& \multicolumn{1}{c}{$\log\left(\frac{1+\rho}{1-\rho}\right)$}
& \multicolumn{1}{c}{$\beta_0$}
& \multicolumn{1}{c}{$\beta_1$}
& \multicolumn{1}{c}{$\beta_2$}
& \multicolumn{1}{c}{$\beta_3$} & \multicolumn{1}{c|@{}}{}\\
& True values & 0.693 & 1.386 & 2.944 & -2.500 & 1.000 & -1.000 &
-0.500 &\\
& Uncorrected INLA & 0.700 & 1.530 & 3.133 & -2.543 & 1.027 & -0.947 &
-0.460 &\\
& Corrected INLA & 0.662 & 1.471 & 3.062 & -2.553 & 1.031 & -0.954 &
-0.465 &\\
& MCMC & 0.605 & 1.466 & 3.224 & -2.569 & 1.037 & -0.959 & -0.448 &\\
\multicolumn{10}{|c|@{}}{}\\
\multicolumn{10}{|c|@{}}{\textbf{Comparison between INLA and MCMC,
(E(INLA)-E(MCMC))/sd(MCMC):}}\\
& & \multicolumn{1}{c}{$\log(\sigma_0^{-2})$}
& \multicolumn{1}{c}{$\log(\sigma_1^{-2})$}
& \multicolumn{1}{c}{$\log\left(\frac{1+\rho}{1-\rho}\right)$}
& \multicolumn{1}{c}{$\beta_0$}
& \multicolumn{1}{c}{$\beta_1$}
& \multicolumn{1}{c}{$\beta_2$}
& \multicolumn{1}{c}{$\beta_3$} & \multicolumn{1}{c|@{}}{}\\
& Uncorrected INLA & 0.195 & 0.130 & -0.107 & 0.110 & -0.080 & 0.034 &
-0.059 &\\
& Corrected INLA & 0.113 & -0.007 & -0.183 & 0.066 & -0.049 & 0.013 &
-0.087 &\\
\multicolumn{10}{|c|@{}}{}\\
\multicolumn{10}{|c|@{}}{\textbf{Ratio of variances,
Var(INLA)/Var(MCMC):}}\\
& & \multicolumn{1}{c}{$\log(\sigma_0^{-2})$}
& \multicolumn{1}{c}{$\log(\sigma_1^{-2})$}
& \multicolumn{1}{c}{$\log\left(\frac{1+\rho}{1-\rho}\right)$}
& \multicolumn{1}{c}{$\beta_0$}
& \multicolumn{1}{c}{$\beta_1$}
& \multicolumn{1}{c}{$\beta_2$}
& \multicolumn{1}{c}{$\beta_3$} & \multicolumn{1}{c|@{}}{}\\
& Uncorrected INLA & 0.958 & 1.023 & 0.943 & 0.905 & 0.921 & 0.920 &
0.950 &\\
& Corrected INLA & 0.989 & 1.050 & 1.087 & 0.925 & 0.951 & 0.932 &
0.973 &\\
\multicolumn{10}{|c|@{}}{}\\
\multicolumn{10}{|c|@{}}{\textbf{Average coverage of 95\% intervals from
INLA over MCMC samples:}}\\
& & \multicolumn{1}{c}{$\log(\sigma_0^{-2})$}
& \multicolumn{1}{c}{$\log(\sigma_1^{-2})$}
& \multicolumn{1}{c}{$\log\left(\frac{1+\rho}{1-\rho}\right)$}
& \multicolumn{1}{c}{$\beta_0$}
& \multicolumn{1}{c}{$\beta_1$}
& \multicolumn{1}{c}{$\beta_2$}
& \multicolumn{1}{c}{$\beta_3$} & \multicolumn{1}{c|@{}}{}\\
& Uncorrected INLA & 93.4\% & 94.7\% & 93.7\% & 93.9\% & 94.1\% &
94.0\% & 94.3\% &\\
& Corrected INLA & 94.4\% & 95.4\% & 94.6\% & 94.1\% & 94.4\% & 94.2\%
& 94.5\% &\\
\hline
\end{tabular*}
\end{table}

%
\begin{table}[h!]
\tabcolsep=4pt
\caption{Results from simulation study~\eqref{eq:fongbivar} with
Var$(b_{0i}) = 3$, Var$(b_{1i}) = 0.5$ and $\rho=0$}
\label{table6}
\renewcommand*{\arraystretch}{1.1}
\begin{tabular*}{\textwidth}{@{\extracolsep{\fill}}|llcccccccc|@{}}
\hline
\multicolumn{10}{|c|@{}}{\textbf{Averages of posterior means:}}\\
& & \multicolumn{1}{c}{$\log(\sigma_0^{-2})$}
& \multicolumn{1}{c}{$\log(\sigma_1^{-2})$}
& \multicolumn{1}{c}{$\log\left(\frac{1+\rho}{1-\rho}\right)$}
& \multicolumn{1}{c}{$\beta_0$}
& \multicolumn{1}{c}{$\beta_1$}
& \multicolumn{1}{c}{$\beta_2$}
& \multicolumn{1}{c}{$\beta_3$} & \multicolumn{1}{c|@{}}{}\\
&True values & -1.099 & 0.693 & 0.000 & -2.500 & 1.000 & -1.000 &
-0.500 &\\
& Uncorrected INLA & -0.749 & 0.962 & -0.243 & -2.274 & 0.956 & -0.661
& -0.584 &\\
& Corrected INLA & -0.809 & 0.850 & -0.294 & -2.318 & 0.978 & -0.672 &
-0.592 &\\
& MCMC & -0.844 & 0.867 & -0.250 & -2.306 & 0.971 & -0.652 & -0.593 &\\
\multicolumn{10}{|c|@{}}{}\\
\multicolumn{10}{|c|@{}}{\textbf{Comparison between INLA and MCMC,
(E(INLA)-E(MCMC))/sd(MCMC):}}\\
& & \multicolumn{1}{c}{$\log(\sigma_0^{-2})$}
& \multicolumn{1}{c}{$\log(\sigma_1^{-2})$}
& \multicolumn{1}{c}{$\log\left(\frac{1+\rho}{1-\rho}\right)$}
& \multicolumn{1}{c}{$\beta_0$}
& \multicolumn{1}{c}{$\beta_1$}
& \multicolumn{1}{c}{$\beta_2$}
& \multicolumn{1}{c}{$\beta_3$} & \multicolumn{1}{c|@{}}{}\\
&Uncorrected INLA & 0.335 & 0.317 & 0.017 & 0.101 & -0.106 & -0.021 &
0.049 &\\
& Corrected INLA & 0.126 & -0.058 & -0.106 & -0.039 & 0.047 & -0.050 &
0.003 &\\
\multicolumn{10}{|c|@{}}{}\\
\multicolumn{10}{|c|@{}}{\textbf{Ratio of variances,
Var(INLA)/Var(MCMC):}}\\
& & \multicolumn{1}{c}{$\log(\sigma_0^{-2})$}
& \multicolumn{1}{c}{$\log(\sigma_1^{-2})$}
& \multicolumn{1}{c}{$\log\left(\frac{1+\rho}{1-\rho}\right)$}
& \multicolumn{1}{c}{$\beta_0$}
& \multicolumn{1}{c}{$\beta_1$}
& \multicolumn{1}{c}{$\beta_2$}
& \multicolumn{1}{c}{$\beta_3$} & \multicolumn{1}{c|@{}}{}\\
&Uncorrected INLA & 0.953 & 0.986 & 0.984 & 0.902 & 0.908 & 0.886 &
0.903 &\\
& Corrected INLA & 0.951 & 0.969 & 0.950 & 0.950 & 0.978 & 0.933 &
0.976 &\\
\multicolumn{10}{|c|@{}}{}\\
\multicolumn{10}{|c|@{}}{\textbf{Average coverage of 95\% intervals from
INLA over MCMC samples:}}\\
& & \multicolumn{1}{c}{$\log(\sigma_0^{-2})$}
& \multicolumn{1}{c}{$\log(\sigma_1^{-2})$}
& \multicolumn{1}{c}{$\log\left(\frac{1+\rho}{1-\rho}\right)$}
& \multicolumn{1}{c}{$\beta_0$}
& \multicolumn{1}{c}{$\beta_1$}
& \multicolumn{1}{c}{$\beta_2$}
& \multicolumn{1}{c}{$\beta_3$} & \multicolumn{1}{c|@{}}{}\\
&Uncorrected INLA & 92.9\% & 93.2\% & 94.8\% & 93.9\% & 93.9\% & 93.6\%
& 93.8\% &\\
& Corrected INLA & 94.2\% & 94.7\% & 94.3\% & 94.3\% & 94.6\% & 94.2\%
& 94.7\% &\\
\hline
\end{tabular*}
\end{table}

%
\begin{table}[t!]
\tabcolsep=4pt
\caption{Results from simulation study~\eqref{eq:fongbivar} with
Var$(b_{0i}) = 3$, Var$(b_{1i}) = 0.5$ and $\rho=0.5$}
\label{table7}
\renewcommand*{\arraystretch}{1.1}
\begin{tabular*}{\textwidth}{@{\extracolsep{\fill}}|llcccccccc|@{}}
\hline
\multicolumn{10}{|c|@{}}{\textbf{Averages of posterior means:}}\\
& & \multicolumn{1}{c}{$\log(\sigma_0^{-2})$}
& \multicolumn{1}{c}{$\log(\sigma_1^{-2})$}
& \multicolumn{1}{c}{$\log\left(\frac{1+\rho}{1-\rho}\right)$}
& \multicolumn{1}{c}{$\beta_0$}
& \multicolumn{1}{c}{$\beta_1$}
& \multicolumn{1}{c}{$\beta_2$}
& \multicolumn{1}{c}{$\beta_3$} & \multicolumn{1}{c|@{}}{}\\
& True values & -1.099 & 0.693 & 1.099 & -2.500 & 1.000 & -1.000 &
-0.500 &\\
& Uncorrected INLA & -0.714 & 1.289 & 1.778 & -2.151 & 1.039 & -1.013
& -0.559 &\\
& Corrected INLA & -0.816 & 1.106 & 1.348 & -2.229 & 1.080 & -1.041 &
-0.570 &\\
& MCMC & -0.816 & 1.172 & 1.643 & -2.166 & 1.048 & -1.003 & -0.557 &\\
\multicolumn{10}{|c|@{}}{}\\
\multicolumn{10}{|c|@{}}{\textbf{Comparison between INLA and MCMC,
(E(INLA)-E(MCMC))/sd(MCMC):}}\\
& & \multicolumn{1}{c}{$\log(\sigma_0^{-2})$}
& \multicolumn{1}{c}{$\log(\sigma_1^{-2})$}
& \multicolumn{1}{c}{$\log\left(\frac{1+\rho}{1-\rho}\right)$}
& \multicolumn{1}{c}{$\beta_0$}
& \multicolumn{1}{c}{$\beta_1$}
& \multicolumn{1}{c}{$\beta_2$}
& \multicolumn{1}{c}{$\beta_3$} & \multicolumn{1}{c|@{}}{}\\
&Uncorrected INLA & 0.336 & 0.273 & 0.141 & 0.045 & -0.061 & -0.022 &
-0.010 &\\
& Corrected INLA & 0.002 & -0.170 & -0.306 & -0.203 & 0.226 & -0.091 &
-0.068 &\\
\multicolumn{10}{|c|@{}}{}\\
\multicolumn{10}{|c|@{}}{\textbf{Ratio of variances,
Var(INLA)/Var(MCMC):}}\\
& & \multicolumn{1}{c}{$\log(\sigma_0^{-2})$}
& \multicolumn{1}{c}{$\log(\sigma_1^{-2})$}
& \multicolumn{1}{c}{$\log\left(\frac{1+\rho}{1-\rho}\right)$}
& \multicolumn{1}{c}{$\beta_0$}
& \multicolumn{1}{c}{$\beta_1$}
& \multicolumn{1}{c}{$\beta_2$}
& \multicolumn{1}{c}{$\beta_3$} & \multicolumn{1}{c|@{}}{}\\
& Uncorrected INLA & 0.954 & 1.030 & 0.977 & 0.892 & 0.905 & 0.901 &
0.934 &\\
& Corrected INLA & 0.934 & 0.964 & 0.761 & 0.984 & 1.031 & 0.976 &
1.041 &\\
\multicolumn{10}{|c|@{}}{}\\
\multicolumn{10}{|c|@{}}{\textbf{Average coverage of 95\% intervals from
INLA over MCMC samples:}}\\
& & \multicolumn{1}{c}{$\log(\sigma_0^{-2})$}
& \multicolumn{1}{c}{$\log(\sigma_1^{-2})$}
& \multicolumn{1}{c}{$\log\left(\frac{1+\rho}{1-\rho}\right)$}
& \multicolumn{1}{c}{$\beta_0$}
& \multicolumn{1}{c}{$\beta_1$}
& \multicolumn{1}{c}{$\beta_2$}
& \multicolumn{1}{c}{$\beta_3$} & \multicolumn{1}{c|@{}}{}\\
& Uncorrected INLA & 92.8\% & 93.5\% & 93.7\% & 93.5\% & 93.6\% &
93.7\% & 94.2\% & \\
& Corrected INLA & 93.3\% & 94.2\% & 90.5\% & 93.7\% & 94.2\% & 94.6\%
& 95.3\% &\\
\hline
\end{tabular*}
\end{table}

%
\begin{table}[h!]
\tabcolsep=4pt
\caption{Results from simulation study~\eqref{eq:fongbivar} with
Var$(b_{0i}) = 3$, Var$(b_{1i}) = 0.5$ and $\rho=0.9$}
\label{table8}
\renewcommand*{\arraystretch}{1.1}
\begin{tabular*}{\textwidth}{@{\extracolsep{\fill}}|llcccccccc|@{}}
\hline
\multicolumn{10}{|c|@{}}{\textbf{Averages of posterior means:}}\\
& & \multicolumn{1}{c}{$\log(\sigma_0^{-2})$}
& \multicolumn{1}{c}{$\log(\sigma_1^{-2})$}
& \multicolumn{1}{c}{$\log\left(\frac{1+\rho}{1-\rho}\right)$}
& \multicolumn{1}{c}{$\beta_0$}
& \multicolumn{1}{c}{$\beta_1$}
& \multicolumn{1}{c}{$\beta_2$}
& \multicolumn{1}{c}{$\beta_3$} & \multicolumn{1}{c|@{}}{}\\
&True values & -1.099 & 0.693 & 2.944 & -2.500 & 1.000 & -1.000 &
-0.500 &\\
& Uncorrected INLA & -1.087 & 1.025 & 4.483 & -2.672 & 1.015 & -0.876
& -0.590 &\\
& Corrected INLA & -1.160 & 0.940 & 4.466 & -2.699 & 1.018 & -0.902 &
-0.606 &\\
& MCMC & -1.169 & 0.941 & 4.537 & -2.703 & 1.039 & -0.845 & -0.572 &\\
\multicolumn{10}{|c|@{}}{}\\
\multicolumn{10}{|c|@{}}{\textbf{Comparison between INLA and MCMC,
(E(INLA)-E(MCMC))/sd(MCMC):}}\\
& & \multicolumn{1}{c}{$\log(\sigma_0^{-2})$}
& \multicolumn{1}{c}{$\log(\sigma_1^{-2})$}
& \multicolumn{1}{c}{$\log\left(\frac{1+\rho}{1-\rho}\right)$}
& \multicolumn{1}{c}{$\beta_0$}
& \multicolumn{1}{c}{$\beta_1$}
& \multicolumn{1}{c}{$\beta_2$}
& \multicolumn{1}{c}{$\beta_3$} & \multicolumn{1}{c|@{}}{}\\
& Uncorrected INLA & 0.288 & 0.192 & -0.063 & 0.083 & -0.150 & -0.061 &
-0.079 &\\
& Corrected INLA & 0.032 & -0.008 & -0.075 & 0.011 & -0.130 & -0.111 &
-0.149 &\\
\multicolumn{10}{|c|@{}}{}\\
\multicolumn{10}{|c|@{}}{\textbf{Ratio of variances,
Var(INLA)/Var(MCMC):}}\\
& & \multicolumn{1}{c}{$\log(\sigma_0^{-2})$}
& \multicolumn{1}{c}{$\log(\sigma_1^{-2})$}
& \multicolumn{1}{c}{$\log\left(\frac{1+\rho}{1-\rho}\right)$}
& \multicolumn{1}{c}{$\beta_0$}
& \multicolumn{1}{c}{$\beta_1$}
& \multicolumn{1}{c}{$\beta_2$}
& \multicolumn{1}{c}{$\beta_3$} & \multicolumn{1}{c|@{}}{}\\
& Uncorrected INLA & 0.938 & 0.961 & 0.883 & 0.883 & 0.897 & 0.887 &
0.932 &\\
& Corrected INLA & 0.978 & 1.019 & 1.060 & 0.933 & 0.943 & 0.937 &
0.986 &\\
\multicolumn{10}{|c|@{}}{}\\
\multicolumn{10}{|c|@{}}{\textbf{Average coverage of 95\% intervals from
INLA over MCMC samples:}}\\
& & \multicolumn{1}{c}{$\log(\sigma_0^{-2})$}
& \multicolumn{1}{c}{$\log(\sigma_1^{-2})$}
& \multicolumn{1}{c}{$\log\left(\frac{1+\rho}{1-\rho}\right)$}
& \multicolumn{1}{c}{$\beta_0$}
& \multicolumn{1}{c}{$\beta_1$}
& \multicolumn{1}{c}{$\beta_2$}
& \multicolumn{1}{c}{$\beta_3$} & \multicolumn{1}{c|@{}}{}\\
& Uncorrected INLA & 92.1\% & 92.6\% & 91.4\% & 93.5\% & 93.5\% &
93.5\% & 93.9\% & \\
& Corrected INLA & 93.5\% & 93.9\% & 93.3\% & 94.1\% & 94.1\% & 94.0\%
& 94.3\% & \\
\hline
\end{tabular*}
\end{table}

\subsection{Model with very small number of observations per cluster}
\label{sec:model-with-very}

In the simulation study described in Section~\ref{sec:simulation-study}
we followed~\citet{fong2010} and used $n_i=7$ observations per cluster.
As suggested by a reviewer, we here consider the effect of having an
even smaller value of $n_i$. We only show the results for the most
extreme possible case, which is $n_i=2$. Using the close to
non-informative prior settings of model~\eqref{eq:07} in Section~\ref
{sec:simulation-study} ($N(0,1000)$ priors for the $\beta_i$ and a
Gamma$(0.5,0.0164)$ prior for $\sigma^{-2}$), the case with $n_i=7$ is
already quite difficult. Using the settings described in Section~\ref
{sec:simulation-study}, the simulated data are relatively
low-informative, making stable and reliable inference non-trivial.

In order to study the even more extreme case of $n_i=2$, more
informative priors are needed, otherwise both MCMC and INLA will fail.
For non-informative (or very weakly informative) priors the model is
just too close to being singular. Therefore, to study the case of
$n_i=2$, we use the following priors: $N(0,1)$ for the $\beta_i$, and
also a $N(0,1)$ prior for the log precision, $\log(\sigma^{-2})$. We
used sampling times $t=(t_1,t_2)'=(-3,3)'$, and 200 simulated data
sets. One million MCMC samples (after a burn-in of 100,000) were used
for each data set.

The results are shown in Table~\ref{table9}. The correction seems to
work well, giving improved estimates in nearly all cases. Note in
particular that the 95\% coverage is uniformly improved, and that all
the coverage values are between 93.7\% and 96.2\% when using the correction.
\vfill

%
\begin{table}[h!]
\tabcolsep=4pt
\caption{Results from simulation study with $n_i=2$ observations per cluster}
\label{table9}
\renewcommand*{\arraystretch}{1.1}
\begin{tabular*}{\textwidth}{@{\extracolsep{4in minus 4in}}|llrrrrrrrr|@{}}
\hline
\multicolumn{10}{|c|@{}}{\textbf{Averages of posterior means:}}\\
& & \multicolumn{1}{c}{$\sigma_0^2$}
& \multicolumn{1}{c}{$\sigma_0$}
& \multicolumn{1}{c}{$\log\sigma_0^{-2}$}
& \multicolumn{1}{c}{$\beta_0$}
& \multicolumn{1}{c}{$\beta_1$}
& \multicolumn{1}{c}{$\beta_2$}
& \multicolumn{1}{c}{$\beta_3$} & \multicolumn{1}{c|@{}}{}\\
&True values & 1.000 & 1.000 & 0.000 & -2.500 & 1.000 & -1.000 & -0.500
&\\
& Uncorrected INLA & 0.667 & 0.762 & 0.689 & -1.990 & 0.837 & -1.178 &
-0.423 &\\
& Corrected INLA & 1.104 & 0.933 & 0.358 & -2.032 & 0.860 & -1.219 &
-0.442 &\\
& MCMC & 0.956 & 0.899 & 0.392 & -2.114 & 0.891 & -1.230 & -0.427 &\\
\multicolumn{10}{|c|@{}}{}\\
\multicolumn{10}{|c|@{}}{\textbf{Comparison between INLA and MCMC,
(E(INLA)-E(MCMC))/sd(MCMC):}}\\
& & \multicolumn{1}{c}{$\sigma_0^2$}
& \multicolumn{1}{c}{$\sigma_0$}
& \multicolumn{1}{c}{$\log\sigma_0^{-2}$}
& \multicolumn{1}{c}{$\beta_0$}
& \multicolumn{1}{c}{$\beta_1$}
& \multicolumn{1}{c}{$\beta_2$}
& \multicolumn{1}{c}{$\beta_3$} & \multicolumn{1}{c|@{}}{}\\
& Uncorrected INLA & -0.337 & -0.362 & 0.355 & 0.250 & -0.295 & 0.079 &
0.020 &\\
& Corrected INLA & 0.149 & 0.085 & -0.041 & 0.163 & -0.168 & 0.013 &
-0.059& \\
\multicolumn{10}{|c|@{}}{}\\
\multicolumn{10}{|c|@{}}{\textbf{Ratio of variances,
Var(INLA)/Var(MCMC):}}\\
& & \multicolumn{1}{c}{$\sigma_0^2$}
& \multicolumn{1}{c}{$\sigma_0$}
& \multicolumn{1}{c}{$\log\sigma_0^{-2}$}
& \multicolumn{1}{c}{$\beta_0$}
& \multicolumn{1}{c}{$\beta_1$}
& \multicolumn{1}{c}{$\beta_2$}
& \multicolumn{1}{c}{$\beta_3$} & \multicolumn{1}{c|@{}}{}\\
&Uncorrected INLA & 0.393 & 0.592 & 0.841 & 0.876 & 0.823 & 0.902 &
0.873 &\\
& Corrected INLA & 2.030 & 1.360 & 1.127 & 0.920 & 0.896 & 0.933 &
0.907& \\
\multicolumn{10}{|c|@{}}{}\\
\multicolumn{10}{|c|@{}}{\textbf{Average coverage of 95\% intervals from
INLA over MCMC samples:}}\\
& & \multicolumn{1}{c}{$\sigma_0^2$}
& \multicolumn{1}{c}{$\sigma_0$}
& \multicolumn{1}{c}{$\log\sigma_0^{-2}$}
& \multicolumn{1}{c}{$\beta_0$}
& \multicolumn{1}{c}{$\beta_1$}
& \multicolumn{1}{c}{$\beta_2$}
& \multicolumn{1}{c}{$\beta_3$} & \multicolumn{1}{c|@{}}{}\\
& Uncorrected INLA & 89.2\% & 89.2\% & 89.5\% & 93.5\% & 92.6\% &
93.7\% & 93.3\%& \\
& Corrected INLA & 96.2\% & 95.9\% & 96.0\% & 94.2\% & 93.8\% & 94.1\%
& 93.7\%& \\
\hline
\end{tabular*}
\end{table}

\subsection{Simulations with misspecified model}
\label{sec:simul-with-missp}

As suggested by a reviewer, we here study the effect of the case of
estimation from a misspecified model: We simulate data from the
model~\eqref{eq:08} and estimate using model~\eqref{eq:07} (with prior
settings as in Section~\ref{sec:simulation-study}). As before, we use
extremely long MCMC chains (one million iterations after discarding
100,000 iterations) as the ``gold standard''. We simulated 200 data
sets from each of the six configurations described in Appendix~\ref
{sec:model-with-two}. The results are shown in Tables~\ref
{table10}--\ref{table15}. Again, the correction improves the results in
nearly all cases, so it does not seem like using a misspecified model
presents any particular problems for the INLA correction.
%
\begin{table}[h!]
\tabcolsep=0pt
\caption{Results from simulation study with incorrectly specified
model; configuration with Var$(b_{0i}) = 0.5$, Var$(b_{1i}) = 0.25$ and
$\rho=0$}
\label{table10}
\renewcommand*{\arraystretch}{1.15}
\begin{tabular*}{\textwidth}{@{\extracolsep{4in minus 4in}}|llrrrrrrrr|@{}}
\hline
\multicolumn{10}{|c|@{}}{\textbf{Averages of posterior means:}}\\
& & \multicolumn{1}{c}{$\sigma_0^2$}
& \multicolumn{1}{c}{$\sigma_0$}
& \multicolumn{1}{c}{$\log\sigma_0^{-2}$}
& \multicolumn{1}{c}{$\beta_0$}
& \multicolumn{1}{c}{$\beta_1$}
& \multicolumn{1}{c}{$\beta_2$}
& \multicolumn{1}{c}{$\beta_3$} & \multicolumn{1}{c|@{}}{}\\
& Uncorrected INLA & 0.424 & 0.536 & 1.778 & -2.482 & 1.068 & -0.833 &
-0.561& \\
& Corrected INLA & 0.591 & 0.640 & 1.440 & -2.541 & 1.093 & -0.847 &
-0.579 &\\
& MCMC & 0.623 & 0.660 & 1.374 & -2.533 & 1.091 & -0.855 & -0.562& \\
\multicolumn{10}{|c|@{}}{}\\
\multicolumn{10}{|c|@{}}{\textbf{Comparison between INLA and MCMC,
(E(INLA)-E(MCMC))/sd(MCMC):}}\\
& & \multicolumn{1}{c}{$\sigma_0^2$}
& \multicolumn{1}{c}{$\sigma_0$}
& \multicolumn{1}{c}{$\log\sigma_0^{-2}$}
& \multicolumn{1}{c}{$\beta_0$}
& \multicolumn{1}{c}{$\beta_1$}
& \multicolumn{1}{c}{$\beta_2$}
& \multicolumn{1}{c}{$\beta_3$} & \multicolumn{1}{c|@{}}{}\\
&Uncorrected INLA & -0.378 & -0.394 & 0.361 & 0.150 & -0.147 & 0.042 &
0.007 &\\
& Corrected INLA & -0.069 & -0.067 & 0.058 & -0.019 & 0.011 & 0.012 &
-0.079 &\\
\multicolumn{10}{|c|@{}}{}\\
\multicolumn{10}{|c|@{}}{\textbf{Ratio of variances,
Var(INLA)/Var(MCMC):}}\\
& & \multicolumn{1}{c}{$\sigma_0^2$}
& \multicolumn{1}{c}{$\sigma_0$}
& \multicolumn{1}{c}{$\log\sigma_0^{-2}$}
& \multicolumn{1}{c}{$\beta_0$}
& \multicolumn{1}{c}{$\beta_1$}
& \multicolumn{1}{c}{$\beta_2$}
& \multicolumn{1}{c}{$\beta_3$} & \multicolumn{1}{c|@{}}{}\\
& Uncorrected INLA & 0.492 & 0.714 & 1.044 & 0.834 & 0.845 & 0.900 &
0.898 &\\
& Corrected INLA & 0.888 & 0.923 & 1.002 & 0.905 & 0.881 & 0.948 &
0.916 &\\
\multicolumn{10}{|c|@{}}{}\\
\multicolumn{10}{|c|@{}}{\textbf{Average coverage of 95\% intervals from
INLA over MCMC samples:}}\\
& & \multicolumn{1}{c}{$\sigma_0^2$}
& \multicolumn{1}{c}{$\sigma_0$}
& \multicolumn{1}{c}{$\log\sigma_0^{-2}$}
& \multicolumn{1}{c}{$\beta_0$}
& \multicolumn{1}{c}{$\beta_1$}
& \multicolumn{1}{c}{$\beta_2$}
& \multicolumn{1}{c}{$\beta_3$} & \multicolumn{1}{c|@{}}{}\\
& Uncorrected INLA & 87.6\% & 87.3\% & 88.4\% & 92.8\% & 92.9\% &
93.7\% & 93.6\% & \\
& Corrected INLA & 93.9\% & 93.4\% & 93.9\% & 93.6\% & 93.3\% & 94.4\%
& 93.8\% & \\
\hline
\end{tabular*}
\end{table}

%
\begin{table}[t!]
\vspace*{6pt}
\tabcolsep=0pt
\caption{Results from simulation study with incorrectly specified
model; configuration with Var$(b_{0i}) = 0.5$, Var$(b_{1i}) = 0.25$ and
$\rho=0.5$}
\label{table11}
\renewcommand*{\arraystretch}{1.15}
\begin{tabular*}{\textwidth}{@{\extracolsep{4in minus 4in}}|llrrrrrrrr|@{}}
\hline
\multicolumn{10}{|c|@{}}{\textbf{Averages of posterior means:}}\\
& & \multicolumn{1}{c}{$\sigma_0^2$}
& \multicolumn{1}{c}{$\sigma_0$}
& \multicolumn{1}{c}{$\log\sigma_0^{-2}$}
& \multicolumn{1}{c}{$\beta_0$}
& \multicolumn{1}{c}{$\beta_1$}
& \multicolumn{1}{c}{$\beta_2$}
& \multicolumn{1}{c}{$\beta_3$} & \multicolumn{1}{c|@{}}{}\\
&Uncorrected INLA & 0.983 & 0.890 & 0.585 & -2.667 & 0.973 & -1.060 &
-0.220& \\
& Corrected INLA & 1.255 & 1.018 & 0.282 & -2.737 & 0.996 & -1.088 &
-0.230 &\\
& MCMC & 1.362 & 1.066 & 0.172 & -2.746 & 1.004 & -1.111 & -0.214 &\\
\multicolumn{10}{|c|@{}}{}\\
\multicolumn{10}{|c|@{}}{\textbf{Comparison between INLA and MCMC,
(E(INLA)-E(MCMC))/sd(MCMC):}}\\
& & \multicolumn{1}{c}{$\sigma_0^2$}
& \multicolumn{1}{c}{$\sigma_0$}
& \multicolumn{1}{c}{$\log\sigma_0^{-2}$}
& \multicolumn{1}{c}{$\beta_0$}
& \multicolumn{1}{c}{$\beta_1$}
& \multicolumn{1}{c}{$\beta_2$}
& \multicolumn{1}{c}{$\beta_3$} & \multicolumn{1}{c|@{}}{}\\
&Uncorrected INLA & -0.498 & -0.541 & 0.544 & 0.202 & -0.202 & 0.084 &
-0.024 &\\
& Corrected INLA & -0.147 & -0.150 & 0.140 & 0.017 & -0.050 & 0.038 &
-0.069 &\\
\multicolumn{10}{|c|@{}}{}\\
\multicolumn{10}{|c|@{}}{\textbf{Ratio of variances,
Var(INLA)/Var(MCMC):}}\\
& & \multicolumn{1}{c}{$\sigma_0^2$}
& \multicolumn{1}{c}{$\sigma_0$}
& \multicolumn{1}{c}{$\log\sigma_0^{-2}$}
& \multicolumn{1}{c}{$\beta_0$}
& \multicolumn{1}{c}{$\beta_1$}
& \multicolumn{1}{c}{$\beta_2$}
& \multicolumn{1}{c}{$\beta_3$} & \multicolumn{1}{c|@{}}{}\\
&Uncorrected INLA & 0.592 & 0.854 & 1.318 & 0.780 & 0.812 & 0.850 &
0.869 &\\
& Corrected INLA & 0.851 & 0.943 & 1.059 & 0.851 & 0.842 & 0.905 &
0.888 &\\
\multicolumn{10}{|c|@{}}{}\\
\multicolumn{10}{|c|@{}}{\textbf{Average coverage of 95\% intervals from
INLA over MCMC samples:}}\\
& & \multicolumn{1}{c}{$\sigma_0^2$}
& \multicolumn{1}{c}{$\sigma_0$}
& \multicolumn{1}{c}{$\log\sigma_0^{-2}$}
& \multicolumn{1}{c}{$\beta_0$}
& \multicolumn{1}{c}{$\beta_1$}
& \multicolumn{1}{c}{$\beta_2$}
& \multicolumn{1}{c}{$\beta_3$} & \multicolumn{1}{c|@{}}{}\\
&Uncorrected INLA & 89.3\% & 89.1\% & 89.6\% & 91.8\% & 92.2\% & 93.0\%
& 93.3\% &\\
& Corrected INLA & 94.0\% & 93.8\% & 94.1\% & 93.1\% & 92.9\% & 93.8\%
& 93.6\% &\\
\hline
\end{tabular*}
\end{table}

%
\begin{table}[h!]
\tabcolsep=0pt
\caption{Results from simulation study with incorrectly specified
model; configuration with Var$(b_{0i}) = 0.5$, Var$(b_{1i}) = 0.25$ and
$\rho=0.9$}
\label{table12}
\renewcommand*{\arraystretch}{1.15}
\begin{tabular*}{\textwidth}{@{\extracolsep{4in minus 4in}}|llrrrrrrrr|@{}}
\hline
\multicolumn{10}{|c|@{}}{\textbf{Averages of posterior means:}}\\
& & \multicolumn{1}{c}{$\sigma_0^2$}
& \multicolumn{1}{c}{$\sigma_0$}
& \multicolumn{1}{c}{$\log\sigma_0^{-2}$}
& \multicolumn{1}{c}{$\beta_0$}
& \multicolumn{1}{c}{$\beta_1$}
& \multicolumn{1}{c}{$\beta_2$}
& \multicolumn{1}{c}{$\beta_3$} & \multicolumn{1}{c|@{}}{}\\
& Uncorrected INLA & 1.071 & 0.937 & 0.451 & -2.621 & 1.097 & -1.243 &
-0.478 &\\
& Corrected INLA & 1.447 & 1.104 & 0.073 & -2.713 & 1.133 & -1.287 &
-0.501 &\\
& MCMC & 1.471 & 1.118 & 0.037 & -2.698 & 1.132 & -1.295 & -0.478& \\
\multicolumn{10}{|c|@{}}{}\\
\multicolumn{10}{|c|@{}}{\textbf{Comparison between INLA and MCMC,
(E(INLA)-E(MCMC))/sd(MCMC):}}\\
& & \multicolumn{1}{c}{$\sigma_0^2$}
& \multicolumn{1}{c}{$\sigma_0$}
& \multicolumn{1}{c}{$\log\sigma_0^{-2}$}
& \multicolumn{1}{c}{$\beta_0$}
& \multicolumn{1}{c}{$\beta_1$}
& \multicolumn{1}{c}{$\beta_2$}
& \multicolumn{1}{c}{$\beta_3$} & \multicolumn{1}{c|@{}}{}\\
&Uncorrected INLA & -0.496 & -0.538 & 0.542 & 0.192 & -0.208 & 0.081 &
0.004 &\\
& Corrected INLA & -0.039 & -0.042 & 0.041 & -0.046 & 0.015 & 0.008 &
-0.099 &\\
\multicolumn{10}{|c|@{}}{}\\
\multicolumn{10}{|c|@{}}{\textbf{Ratio of variances,
Var(INLA)/Var(MCMC):}}\\
& & \multicolumn{1}{c}{$\sigma_0^2$}
& \multicolumn{1}{c}{$\sigma_0$}
& \multicolumn{1}{c}{$\log\sigma_0^{-2}$}
& \multicolumn{1}{c}{$\beta_0$}
& \multicolumn{1}{c}{$\beta_1$}
& \multicolumn{1}{c}{$\beta_2$}
& \multicolumn{1}{c}{$\beta_3$} & \multicolumn{1}{c|@{}}{}\\
&Uncorrected INLA & 0.596 & 0.854 & 1.308 & 0.775 & 0.783 & 0.842 &
0.855 &\\
& Corrected INLA & 0.959 & 0.989 & 1.027 & 0.871 & 0.830 & 0.918 &
0.881 &\\
\multicolumn{10}{|c|@{}}{}\\
\multicolumn{10}{|c|@{}}{\textbf{Average coverage of 95\% intervals from
INLA over MCMC samples:}}\\
& & \multicolumn{1}{c}{$\sigma_0^2$}
& \multicolumn{1}{c}{$\sigma_0$}
& \multicolumn{1}{c}{$\log\sigma_0^{-2}$}
& \multicolumn{1}{c}{$\beta_0$}
& \multicolumn{1}{c}{$\beta_1$}
& \multicolumn{1}{c}{$\beta_2$}
& \multicolumn{1}{c}{$\beta_3$} & \multicolumn{1}{c|@{}}{}\\
&Uncorrected INLA & 89.5\% & 89.4\% & 89.8\% & 91.7\% & 91.6\% & 92.9\%
& 93.0\% &\\
& Corrected INLA & 95.0\% & 94.8\% & 95.0\% & 93.2\% & 92.6\% & 94.0\%
& 93.4\% &\\
\hline
\end{tabular*}
\end{table}

%
\begin{table}[t!]
\vspace*{6pt}
\tabcolsep=0pt
\caption{Results from simulation study with incorrectly specified
model; configuration with Var$(b_{0i}) = 3$, Var$(b_{1i}) = 0.5$ and
$\rho=0$}
\label{table13}
\renewcommand*{\arraystretch}{1.15}
\begin{tabular*}{\textwidth}{@{\extracolsep{4in minus 4in}}|llrrrrrrrr|@{}}
\hline
\multicolumn{10}{|c|@{}}{\textbf{Averages of posterior means:}}\\
& & \multicolumn{1}{c}{$\sigma_0^2$}
& \multicolumn{1}{c}{$\sigma_0$}
& \multicolumn{1}{c}{$\log\sigma_0^{-2}$}
& \multicolumn{1}{c}{$\beta_0$}
& \multicolumn{1}{c}{$\beta_1$}
& \multicolumn{1}{c}{$\beta_2$}
& \multicolumn{1}{c}{$\beta_3$} & \multicolumn{1}{c|@{}}{}\\
&Uncorrected INLA & 2.644 & 1.587 & -0.873 & -2.461 & 0.960 & -0.626 &
-0.486&\\
& Corrected INLA & 2.938 & 1.671 & -0.976 & -2.508 & 0.977 & -0.637 &
-0.497& \\
& MCMC & 3.161 & 1.736 & -1.053 & -2.546 & 0.998 & -0.623 & -0.505& \\
\multicolumn{10}{|c|@{}}{}\\
\multicolumn{10}{|c|@{}}{\textbf{Comparison between INLA and MCMC,
(E(INLA)-E(MCMC))/sd(MCMC):}}\\
& & \multicolumn{1}{c}{$\sigma_0^2$}
& \multicolumn{1}{c}{$\sigma_0$}
& \multicolumn{1}{c}{$\log\sigma_0^{-2}$}
& \multicolumn{1}{c}{$\beta_0$}
& \multicolumn{1}{c}{$\beta_1$}
& \multicolumn{1}{c}{$\beta_2$}
& \multicolumn{1}{c}{$\beta_3$} & \multicolumn{1}{c|@{}}{}\\
&Uncorrected INLA & -0.471 & -0.500 & 0.523 & 0.215 & -0.271 & -0.005 &
0.103 &\\
& Corrected INLA & -0.211 & -0.219 & 0.225 & 0.096 & -0.153 & -0.026 &
0.043 &\\
\multicolumn{10}{|c|@{}}{}\\
\multicolumn{10}{|c|@{}}{\textbf{Ratio of variances,
Var(INLA)/Var(MCMC):}}\\
& & \multicolumn{1}{c}{$\sigma_0^2$}
& \multicolumn{1}{c}{$\sigma_0$}
& \multicolumn{1}{c}{$\log\sigma_0^{-2}$}
& \multicolumn{1}{c}{$\beta_0$}
& \multicolumn{1}{c}{$\beta_1$}
& \multicolumn{1}{c}{$\beta_2$}
& \multicolumn{1}{c}{$\beta_3$} & \multicolumn{1}{c|@{}}{}\\
&Uncorrected INLA & 0.683 & 0.828 & 1.021 & 0.788 & 0.786 & 0.826 &
0.838& \\
& Corrected INLA & 0.848 & 0.921 & 1.018 & 0.843 & 0.811 & 0.880 &
0.858 &\\
\multicolumn{10}{|c|@{}}{}\\
\multicolumn{10}{|c|@{}}{\textbf{Average coverage of 95\% intervals from
INLA over MCMC samples:}}\\
& & \multicolumn{1}{c}{$\sigma_0^2$}
& \multicolumn{1}{c}{$\sigma_0$}
& \multicolumn{1}{c}{$\log\sigma_0^{-2}$}
& \multicolumn{1}{c}{$\beta_0$}
& \multicolumn{1}{c}{$\beta_1$}
& \multicolumn{1}{c}{$\beta_2$}
& \multicolumn{1}{c}{$\beta_3$} & \multicolumn{1}{c|@{}}{}\\
&Uncorrected INLA & 91.1\% & 91.1\% & 91.4\% & 91.8\% & 91.5\% & 92.6\%
& 92.6\% &\\
& Corrected INLA & 94.2\% & 94.2\% & 94.4\% & 93.0\% & 92.4\% & 93.4\%
& 93.1\% & \\
\hline
\end{tabular*}
\end{table}

%
\begin{table}[h!]
\tabcolsep=0pt
\caption{Results from simulation study with incorrectly specified
model; configuration with Var$(b_{0i}) = 3$, Var$(b_{1i}) = 0.5$ and
$\rho=0.5$}
\label{table14}
\renewcommand*{\arraystretch}{1.15}
\begin{tabular*}{\textwidth}{@{\extracolsep{4in minus 4in}}|llrrrrrrrr|@{}}
\hline
\multicolumn{10}{|c|@{}}{\textbf{Averages of posterior means:}}\\
& & \multicolumn{1}{c}{$\sigma_0^2$}
& \multicolumn{1}{c}{$\sigma_0$}
& \multicolumn{1}{c}{$\log\sigma_0^{-2}$}
& \multicolumn{1}{c}{$\beta_0$}
& \multicolumn{1}{c}{$\beta_1$}
& \multicolumn{1}{c}{$\beta_2$}
& \multicolumn{1}{c}{$\beta_3$} & \multicolumn{1}{c|@{}}{}\\
&Uncorrected INLA & 3.246 & 1.749 & -1.056 & -3.025 & 0.936 & -0.642 &
-0.182 &\\
& Corrected INLA & 3.821 & 1.888 & -1.203 & -3.117 & 0.957 & -0.676 &
-0.186 &\\
& MCMC & 3.990 & 1.938 & -1.261 & -3.144 & 0.978 & -0.663 & -0.183 &\\
\multicolumn{10}{|c|@{}}{}\\
\multicolumn{10}{|c|@{}}{\textbf{Comparison between INLA and MCMC,
(E(INLA)-E(MCMC))/sd(MCMC):}}\\
& & \multicolumn{1}{c}{$\sigma_0^2$}
& \multicolumn{1}{c}{$\sigma_0$}
& \multicolumn{1}{c}{$\log\sigma_0^{-2}$}
& \multicolumn{1}{c}{$\beta_0$}
& \multicolumn{1}{c}{$\beta_1$}
& \multicolumn{1}{c}{$\beta_2$}
& \multicolumn{1}{c}{$\beta_3$} & \multicolumn{1}{c|@{}}{}\\
&Uncorrected INLA & -0.506 & -0.541 & 0.570 & 0.251 & -0.277 & 0.032 &
0.009 &\\
& Corrected INLA & -0.144 & -0.154 & 0.161 & 0.063 & -0.139 & -0.017 &
-0.012 &\\
\multicolumn{10}{|c|@{}}{}\\
\multicolumn{10}{|c|@{}}{\textbf{Ratio of variances,
Var(INLA)/Var(MCMC):}}\\
& & \multicolumn{1}{c}{$\sigma_0^2$}
& \multicolumn{1}{c}{$\sigma_0$}
& \multicolumn{1}{c}{$\log\sigma_0^{-2}$}
& \multicolumn{1}{c}{$\beta_0$}
& \multicolumn{1}{c}{$\beta_1$}
& \multicolumn{1}{c}{$\beta_2$}
& \multicolumn{1}{c}{$\beta_3$} & \multicolumn{1}{c|@{}}{}\\
&Uncorrected INLA & 0.643 & 0.802 & 1.020 & 0.761 & 0.793 & 0.806 &
0.830& \\
& Corrected INLA & 0.922 & 0.973 & 1.050 & 0.846 & 0.826 & 0.885 &
0.860& \\
\multicolumn{10}{|c|@{}}{}\\
\multicolumn{10}{|c|@{}}{\textbf{Average coverage of 95\% intervals from
INLA over MCMC samples:}}\\
& & \multicolumn{1}{c}{$\sigma_0^2$}
& \multicolumn{1}{c}{$\sigma_0$}
& \multicolumn{1}{c}{$\log\sigma_0^{-2}$}
& \multicolumn{1}{c}{$\beta_0$}
& \multicolumn{1}{c}{$\beta_1$}
& \multicolumn{1}{c}{$\beta_2$}
& \multicolumn{1}{c}{$\beta_3$} & \multicolumn{1}{c|@{}}{}\\
&Uncorrected INLA & 90.4\% & 90.4\% & 90.7\% & 91.2\% & 91.6\% & 92.3\%
& 92.6\% &\\
& Corrected INLA & 94.8\% & 94.8\% & 95.0\% & 93.0\% & 92.7\% & 93.5\%
& 93.1\% &\\
\hline
\end{tabular*}
\end{table}

%
\begin{table}[t!]
\vspace*{6pt}
\tabcolsep=0pt
\caption{Results from simulation study with incorrectly specified
model; configuration with Var$(b_{0i}) = 3$, Var$(b_{1i}) = 0.5$ and
$\rho=0.9$}
\label{table15}
\renewcommand*{\arraystretch}{1.15}
\begin{tabular*}{\textwidth}{@{\extracolsep{4in minus 4in}}|llrrrrrrrr|@{}}
\hline
\multicolumn{10}{|c|@{}}{\textbf{Averages of posterior means:}}\\
& & \multicolumn{1}{c}{$\sigma_0^2$}
& \multicolumn{1}{c}{$\sigma_0$}
& \multicolumn{1}{c}{$\log\sigma_0^{-2}$}
& \multicolumn{1}{c}{$\beta_0$}
& \multicolumn{1}{c}{$\beta_1$}
& \multicolumn{1}{c}{$\beta_2$}
& \multicolumn{1}{c}{$\beta_3$} & \multicolumn{1}{c|@{}}{}\\
&Uncorrected INLA & 3.622 & 1.854 & -1.180 & -3.170 & 1.160 & -0.564 &
-0.213 &\\
& Corrected INLA & 4.118 & 1.973 & -1.302 & -3.253 & 1.185 & -0.582 &
-0.220 &\\
& MCMC & 4.468 & 2.058 & -1.389 & -3.312 & 1.218 & -0.576 & -0.222 &\\
\multicolumn{10}{|c|@{}}{}\\
\multicolumn{10}{|c|@{}}{\textbf{Comparison between INLA and MCMC,
(E(INLA)-E(MCMC))/sd(MCMC):}}\\
& & \multicolumn{1}{c}{$\sigma_0^2$}
& \multicolumn{1}{c}{$\sigma_0$}
& \multicolumn{1}{c}{$\log\sigma_0^{-2}$}
& \multicolumn{1}{c}{$\beta_0$}
& \multicolumn{1}{c}{$\beta_1$}
& \multicolumn{1}{c}{$\beta_2$}
& \multicolumn{1}{c}{$\beta_3$} & \multicolumn{1}{c|@{}}{}\\
&Uncorrected INLA & -0.526 & -0.564 & 0.595 & 0.276 & -0.326 & 0.017 &
0.037& \\
& Corrected INLA & -0.229 & -0.240 & 0.248 & 0.117 & -0.186 & -0.008 &
0.007& \\
\multicolumn{10}{|c|@{}}{}\\
\multicolumn{10}{|c|@{}}{\textbf{Ratio of variances,
Var(INLA)/Var(MCMC):}}\\
& & \multicolumn{1}{c}{$\sigma_0^2$}
& \multicolumn{1}{c}{$\sigma_0$}
& \multicolumn{1}{c}{$\log\sigma_0^{-2}$}
& \multicolumn{1}{c}{$\beta_0$}
& \multicolumn{1}{c}{$\beta_1$}
& \multicolumn{1}{c}{$\beta_2$}
& \multicolumn{1}{c}{$\beta_3$} & \multicolumn{1}{c|@{}}{}\\
&Uncorrected INLA & 0.641 & 0.802 & 1.021 & 0.751 & 0.755 & 0.797 &
0.804 &\\
& Corrected INLA & 0.844 & 0.925 & 1.033 & 0.817 & 0.788 & 0.860 &
0.832 & \\
\multicolumn{10}{|c|@{}}{}\\
\multicolumn{10}{|c|@{}}{\textbf{Average coverage of 95\% intervals from
INLA over MCMC samples:}}\\
& & \multicolumn{1}{c}{$\sigma_0^2$}
& \multicolumn{1}{c}{$\sigma_0$}
& \multicolumn{1}{c}{$\log\sigma_0^{-2}$}
& \multicolumn{1}{c}{$\beta_0$}
& \multicolumn{1}{c}{$\beta_1$}
& \multicolumn{1}{c}{$\beta_2$}
& \multicolumn{1}{c}{$\beta_3$} & \multicolumn{1}{c|@{}}{}\\
&Uncorrected INLA & 90.1\% & 90.1\% & 90.4\% & 90.9\% & 90.6\% & 92.1\%
& 92.1\% & \\
& Corrected INLA & 94.3\% & 94.2\% & 94.5\% & 92.6\% & 91.9\% & 93.1\%
& 92.7\% & \\
\hline
\end{tabular*}
\end{table}

\end{appendix}

\section*{Acknowledgments} 
We thank Youyi Fong for providing R code relating to the simulation
study described in Section~\ref{sec:simulation-study}. We are also very
grateful to Leonard Held and Rafael Sauter for providing us with a copy
of their unpublished paper along with R code relevant for the analysis
of toenail data described in Section~\ref{sec:toenail}.
We thank Janine Illian, Geir-Arne Fuglstad, Dan Simpson and two
anonymous reviewers for helpful comments that have led to an improved
presentation.
This research was supported by the Norwegian Research Council.\vfill\eject


%

\end{document}